\documentclass[aps,a4paper,fleqn,usenatbib,pra,twocolumn,nopacs,floatfix]{revtex4}
\usepackage{epsfig}
\usepackage{graphicx}
\usepackage{dcolumn}
\usepackage{amsthm,amsmath}
\usepackage{comment}
\usepackage{float}
\usepackage[colorlinks]{hyperref}
\usepackage{xcolor}
\usepackage{ulem}
\usepackage{orcidlink}

\begin{document}

%
\author{$^1$Jyoti\orcidlink{0000-0002-7013-3285}}
\author{$^{2,3}$A. Chakraborty\orcidlink{0000-0001-6255-4584}}
\author{$^{1}$Zhiyang Wang\orcidlink{0000-0002-0931-532X}}
\author{$^{1}$Jia Zhang\orcidlink{0000-0001-8682-8287}}
\author{$^{1,4}$Jingbiao Chen\orcidlink{0000-0001-9802-4577}}
\email{jbchen@pku.edu.cn}
\author{$^{5,6}$Bindiya Arora\orcidlink{0000-0001-7083-034X}}
\author{$^{2}$B. K. Sahoo\orcidlink{0000-0003-4397-7965}}
\email{bijaya@prl.res.in}

\affiliation{$^{1}$Institute of Quantum Electronics, School of Electronics, Peking University, Beijing 100871, P. R. China}
\affiliation{$^{2}$Atomic, Molecular and Optical Physics Division, Physical Research Laboratory, Navrangpura, Ahmedabad-380009, India\\
$^{3}$Indian Institute of Technology Gandhinagar, Palaj, Gandhinagar 382355, India} 
\affiliation{$^4$ Hefei National Laboratory, Hefei 230088, P.R. China}
\affiliation{$^5$Department of Physics, Guru Nanak Dev University, Amritsar, Punjab 143005, India}
\affiliation{$^6$Perimeter Institute for Theoretical Physics, Waterloo, Ontario N2L 2Y5, Canada}

\title{Investigating the $4D_{3/2}|3,\pm2\rangle$--$4D_{5/2}|3,\pm2\rangle$ transition in Nb$^{4+}$ for a THz atomic clock} 


\begin{abstract}
In this work, the $4D_{3/2}|3,\pm2\rangle \rightarrow 4D_{5/2}|3,\pm2\rangle$ transition in the Nb$^{4+}$ ion is identified as a promising candidate for a terahertz (THz) atomic clock, with the transition frequency occurring at 56.0224 THz. This transition is primarily driven by the magnetic dipole decay channel, which can easily be accessed by a laser. We focus on the stable $^{93}$Nb isotope, which has 100\% natural abundance and a nuclear spin of $I=9/2$ for experimental advantage. Our data analysis allows us to estimate potential systematic shifts in the proposed clock system, including those due to blackbody radiation, electric quadrupole, second-order Zeeman, and second-order Doppler {shifts}. {The scheme presented in this study can help suppress the AC Stark and electric quadrupole shifts in the clock frequency measurement.} {All these analyses} suggest that the proposed THz atomic clock using Nb$^{4+}$ could be valuable in both quantum thermometry and frequency metrology. 
\end{abstract}
\maketitle
\date \today

\section{Introduction}

Atomic clocks are often used as the timekeeping devices which define the international standard of time. Recent technological advancement in atomic
physics has improved the precision of these atomic clocks to an extent that these devices only lose a second over several billion years. Atomic 
clocks are recently used in the investigation of various phenomena of fundamental physics including dark matter search \cite{derevianko2014hunting},
variation of fundamental physical constants \cite{rosenband2008frequency}, relativistic geodesy \cite{Mehlstaubler_2018,mcgrew2018atomic}, 
gravitational-wave detection \cite{kolkowitz2016gw,loeb2015using,graham2013new} as well as the interactions describing beyond the Standard Model
particle physics \cite{dzuba2018testing}. Most of the existing atomic clocks operate in both optical and microwave ranges, but lately, the 
operating range of these devices have been expanded to terahertz (THz) range based on the applications of various ingenious modes of THz 
electromagnetic radiations in sensing, spectroscopy and communication~\cite{tonouchi2007cutting} and for the analysis of interstellar matter
\cite{kulesa2011terahertz}.

The THz clock transitions are high frequency transitions of the order of $10^{12}$ Hz, hence they are proved to be highly sensitive to blackbody
radiations {(BBR)}, and can be used in quantum thermometers, especially in remote-sensing satellites~\cite{norrgard2021quantum}. Moreover, THz 
frequency standard also finds its applications in new-generation navigation, communication and sensing systems~\cite{kim2019chip}, commercial THz 
instruments such as detectors, sources and high-resolution THz spectrometers~\cite{yasui2010terahertz}. It is also important to devise THz 
clocks for astronomical instruments, especially astronomical interferometers and new-generation space telescopes for probing unexplored universe
including our galaxy that comprises more than $50\%$ of total luminosity due to THz radiations \cite{consolino2017terahertz,bellini1992tunable}. Furthermore,
the realization of THz frequency standard is evident to understand star formation and decay processes as well as the thermal fluctuations in 
environment due to immense release of greenhouse gases \cite{consolino2017terahertz}.

The proposition of implementation of absolute frequency standards in THz domain was first made by Strumia in 1972 considering fine structure 
transition lines of Mg and Ca metastable triplet states \cite{strumia1972proposal}. Yamamoto \textit{et~al.} first generated tunable THz optical 
clock with 100 GHz to 1 THz frequency range~\cite{yamamoto2002generation}. {Champenois \textit{et~al.} first proposed the THz frequency standard  based on three-photon coherent population trapping in trapped ion clouds~\cite{champenois2007thz}.}  Zhou \textit{et~al.} considered alkaline-earth atoms including Sr, Ca 
and Mg to determine the magic wavelengths of THz clock transitions between their metastable triplet states~\cite{zhou2010magic}. Yu \textit{et~al.}
evaluated AC Stark shifts and magic wavelengths of THz clock transitions in barium \cite{yu2015ac}. Wang \textit{et~al.} probed carbonyl sulphide
(OCS) to realize two different molecular clocks based on sub-THz frequency standard~\cite{wang2018chip}. The miniaturization of atomic clock with 
high-affordability in chip-scale THz~OCS clock has been analyzed by  Kim \textit{et~al.}~\cite{kim2019chip}, whereas Drake \textit{et~al.} 
considered silicon nitride to analyze the performance of THz-rate Kerr microresonator optical clock \cite{drake2019terahertz}. Wang 
\textit{et~al.} proposed chip-scale molecular clocks based on complementary metal-oxide-semiconductor technique with selected OCS transitions
\cite{wang2020terahertz}. A THz molecular clock using vibrational levels of Sr$_2$ has been constructed by Leung \textit{et~al.}
\cite{leung2023tera} and achieved a systematic uncertainty at the level of $10^{-14}$. 

In a recent study, we had proposed a THz clock based on magnetic dipole (M1) transition of the $4D_{3/2}$--$4D_{5/2}$ transition in Zr$^{3+}$ ion
and estimated major systematic shifts associated with this transition \cite{jyoti2023zr}. In this work, we introduce a similar M1 clock 
transition at THz frequency in the Nb$^{4+}$ ion and analyze the corresponding systematic frequency shifts. The results reveal that the Nb$^{4+}$ clock is more
sensitive to magnetic perturbations in comparison to our previously proposed Zr$^{3+}$ clock and other clocks proposed in the THz regime. 
Therefore, the Nb$^{4+}$ ion is particularly advantageous for prospective applications, especially in quantum thermometry, where BBR-induced shifts
can be utilized for calibrating quantum thermometers. The high sensitivity of this clock to magnetic perturbations also suggests its potential to 
enhance sensor sensitivity and more accurate synchronization with instruments operating at THz frequencies, which is critical for various quantum technologies
including quantum metrology.

The manuscript is structured as follows: Sec.~\ref{proposal} presents the detailed proposal for our THz ion clock, Sec.~\ref{2} demonstrates
the method of evaluation of atomic wave functions and matrix elements, {Sec~\ref{energies} demonstrates the estimation of energies of various states and their comparison with  {National Institute of Standards and Technology (NIST)} energies as well as the estimations from different methods,} Sec.~\ref{3} presents electric dipole (E1) and magnetic dipole (M1) polarizabilities used 
for estimating systematic effects. Sec.~\ref{4} discusses the dominant systematic shifts, while the study is concluded in Sec.~\ref{5}. All the 
physical quantities are given in atomic units (a.u.) unless stated otherwise.

\section{Scheme of $^{93}$Nb$^{4+}$ THz clock} \label{proposal}

From our earlier calculations of spectroscopic properties in Nb$^{4+}$ \cite{jyoti2021spectroscopic}, the wavelength of the fine-structure 
splitting $4D_{3/2}$--$4D_{5/2}$ of the ground state in this ion is found to be $\lambda_0=5.355~\mu m$, corresponding to frequency of $56.0224$ THz.
The lifetime of the $4D_{5/2}$ excited state of this transition due to forbidden decay channels is estimated to be approximately 12.65 seconds (s) 
\cite{das2017electron}. These two properties of the $4D_{3/2}$--$4D_{5/2}$ transition in Nb$^{4+}$ fulfil the essential requirements for making 
it a THz clock. $^{93}$Nb has almost 100\% natural abundance \cite{deLaeter2003abundance} with nuclear spin of $I=9/2$, which can be sensitive 
towards detecting magnetic field effects. 

\begin{figure}[t]
\includegraphics[width=9cm]{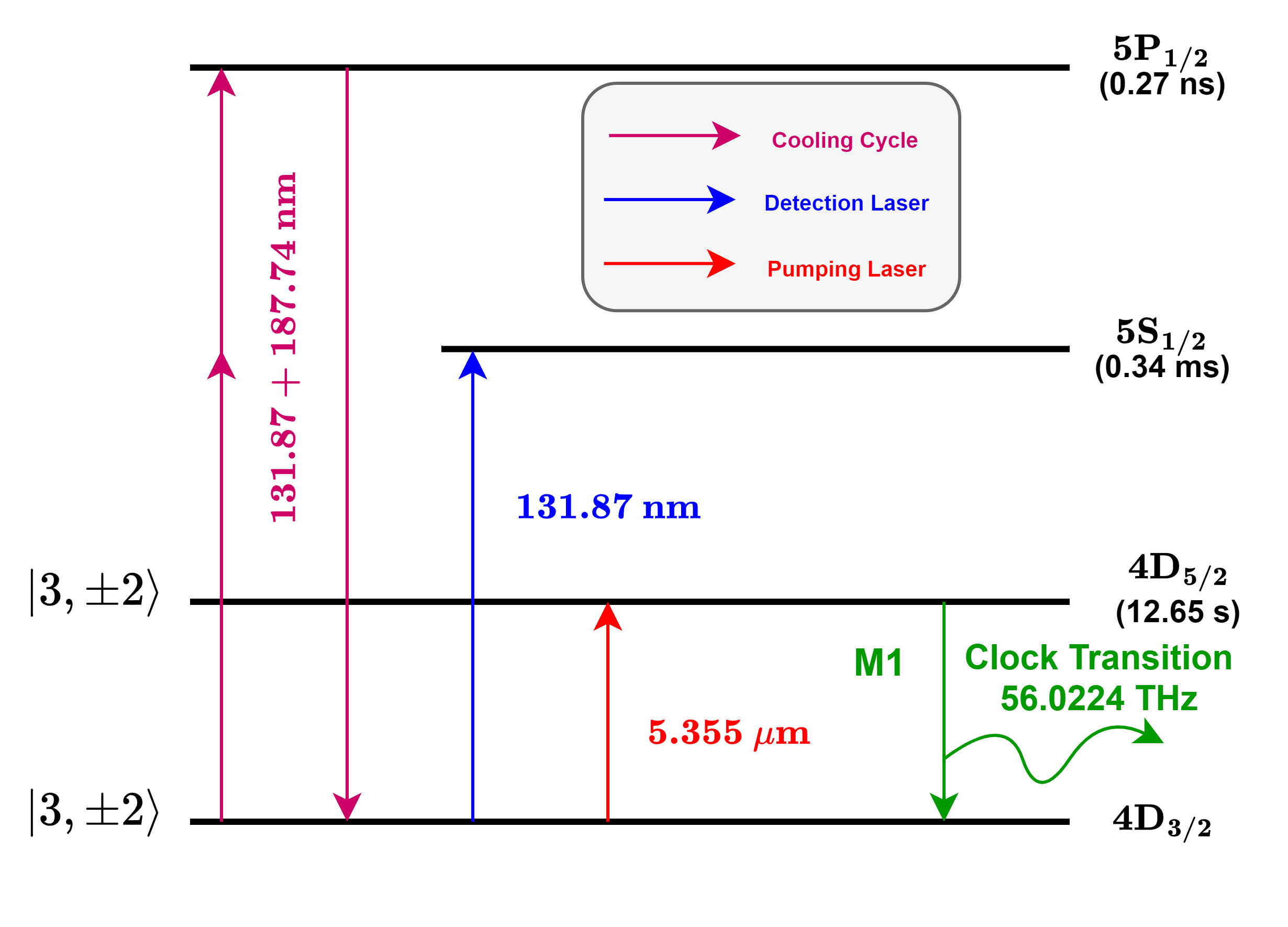}
\caption{\label{figclock}Schematic of clock 
frequency measurement set-up using Nb$^{4+}$ ion. As shown, the $4D_{3/2}[|3,\pm2\rangle]$--$4D_{5/2}[|3,\pm 2\rangle]$ transition is used as both pumping as well as clock transition whereas $4D_{3/2}[|3,\pm2\rangle]$--$5S_{1/2}[|4,0\rangle]$ transition is
used for the detection of population changes at the ground state. The cooling cycle is also explicitly demonstrated in the figure.}
\end{figure}

In a practical experimental setup, trapping $^{93}$Nb$^{4+}$ ion can be achieved using electron beam ion traps \cite{silver1994oxford,nakamura2008compact} or electron cyclotron resonance accelerators~\cite{agnihotri2011ecr}. The clock frequency of the $4D_{3/2}$--$4D_{5/2}$ transition in $^{93}$Nb$^{4+}$ can be measured by trapping this ion simultaneously with either the Mg$^+$ or Al$^+$ ion using quantum logic principle 
as these ions possess similar mass-to-charge ratios as of the Nb$^{4+}$ ion \cite{chou2010al+,hannig2019al}. The schematic diagram of our proposed 
clock is shown in Fig.~\ref{figclock}, in which cooling and interrogation lines are clearly identified. As can be seen, we consider here the hyperfine
levels ($|F, M_F \rangle$) over the atomic states for the actual measurements and the reason for choosing particularly the $|3,\pm 2\rangle$ hyperfine
levels is to minimize systematic effects as discussed later. The sufficiently long lifetime of $12.65$s of the $4D_{5/2}$ state provides a possibility
of long interrogation time. Hence, a desirable amount of atomic population can be accumulated at this state satisfying the population inversion 
criteria between the clock states that is imperative for the clock mechanism to work. Thus, the Nb$^{4+}$ ions can be pumped from the ground 
$4D_{3/2}|3,\pm2\rangle$ state to the clock $4D_{5/2}|3,\pm2\rangle$ state using the already available $5.355 \mu$m pulsed-lasers optimized to an 
intensity required to achieve highest population inversion among the clock states. Besides, these ions can be cooled using two-photon laser cooling 
mechanism through the $4D_{3/2}\rightarrow 5S_{1/2}$ and $ 5S_{1/2} \rightarrow 5P_{1/2}$ transitions corresponding to lasers wavelengths
of $131.87$ nm and $187.74$ nm, respectively. The decay of electrons from the $5P_{1/2}$ state to the $4D_{3/2}$ state with the calculated branching 
ratio of 83\% (high-probability) suggests feasibility of maintaining a recycling process of two laser cooling which is efficient to cool ions, and 
sufficient enough to enhance the coherence and intensity of the clock signal. Further, the population density at the $4D_{3/2}$ clock level can be 
monitored using weak pulses of detection laser of an amplitude of $131.87$ nm implemented between the $4D_{3/2}$ and $5S_{1/2}$ states. The 
implementation of this approach prepares the ion in one of the clock states guaranteeing highly efficient population inversion, thereby, benefitting 
signal-to-noise ratio inherent within the clock. Fundamentally, this clock operation is similar to that of an atomic fountain as seen in rubidium 
atomic fountain clock~\cite{micalizio2021pulsed}.

Experimentally, the implementation of the detection laser can be achieved through high-order harmonic generation techniques, facilitating the 
generation of ultra-short pulses~\cite{Wen_2020,singh2024hhg}. Once a sufficient population of ions is detected at the $4D_{5/2}[|3,\pm2\rangle]$
hyperfine level, the detection laser can be promptly turned off to mitigate resonance effects at the $4D_{3/2}$ state. Subsequently, a stimulated 
emission of clock laser via the M1 decay channel at a frequency of $56.0224$ THz can be observed. It is suggested that both the pumping and detection 
lasers need to be turned off during the observation of clock transition, so that there is no risk of introducing an AC Stark shift and to ensure an 
undisturbed environment. It should also be noted that there is no risk of leakage of ions in this case owing to the proposed compact 
experimental setup and minimal implementation of lasers in the scheme. To achieve high stability and high accuracy in this proposed THz clock scheme,
usage of a feedback loop to control the energy difference between the $4D_{3/2}$ and $4D_{5/2}$ states is recommended. This feedback loop would adjust
the static magnetic field applied to the ion trap for maintaining a stable clock frequency over time~\cite{merkel2019magnetic}. All these 
characteristics of Nb$^{4+}$ ensure that it is a viable candidate for a THz clock.

\section{Method of Evaluation}\label{2}

Precise determination of systematic frequency shifts in a clock transition requires accurate computation of wave functions of the involved states. 
Therefore, we have implemented relativistic coupled cluster (RCC) theory for accurate computation of these wave functions as well as the matrix 
elements of physical operators. Higher-order correlations due to various physical effects accounting for core-polarization and pair-correlation 
effects have also been incorporated through the nonlinear terms of the RCC theory. Although the general formulation and potential applications of
RCC theory can be found in many previous studies including Refs. \cite{blundell1991relativistic,ILYABAEV199482,lindroth1993ab,sahoo2004ab,
nandy2014quadrupole}, still, we provide a brief outline of our employed RCC method below. 

We begin with Dirac-Coulomb Hamiltonian $(H_{DC})$ in our RCC method, which in a.u. is given by
\begin{eqnarray}
H_{DC} &=& \sum_{i=1}^{N_e} \left[  c {\vec \alpha}_D \cdot {\vec p}_i+ (\beta-1) c^2 + V_{n}(r_i) \right ] + \sum_{i>j} \frac{1}{r_{ij}} , \ \ \ \ 
\end{eqnarray}
{where $N_e$ is the number of electrons in the atom, $c$ is the speed of light, ${\vec \alpha}_D$ and $\beta$ are the $4\times 4$ Dirac matrices, $V_n(r)$ is the nuclear potential, and $r_{ij}$ is the inter-electronic distances between electrons located at $r_i$ and $r_j$. Corrections due to Breit and lower-order quantum electrodynamics (QED) are also included to improve accuracy in our calculations~\cite{ginges2016atomic,sahoo2016conform,flambaum2005radiative,li2018cc}. Note that contributions from the negative energy states (NESs) are neglected here, which will not affect our results in the precision of present interest.}

In the RCC theory ansatz, wave function of a many-electron system can be expressed in terms of mean-field (MF) wave function $|\Phi_0 \rangle$ of an 
atomic state and cluster operator $T$ as \cite{vcivzek1969use}
\begin{equation}
|\Psi_0\rangle=e^T|\Phi_0\rangle. \label{eqa}
\end{equation} 
By solving the Dirac-Fock (DF) equation for the closed-shell configurations ($[4p^6]$), we get the MF wave function $|\Phi_0 \rangle$ and then, 
we obtain DF wave function $|\Phi_v\rangle$ of the state with configuration $[4p^6]v$ by adding a suitable valence orbital $v$. In terms of the 
creation operator for the valence electron $a_v^{\dagger}$, it is given as \cite{lindgren2012atomic}
\begin{equation}
|\Phi_{v}\rangle=a_{v}^{\dagger}| \Phi_0 \rangle\, . \label{eqdag}
\end{equation}
 
Wave function of an atomic state with closed-shell electronic configuration and a valence orbital can be expressed in terms of a RCC operator that 
accounts for the excitations of core electrons to virtual orbitals ($T$) and the operator $S_v$ that accounts for the excitation of valence 
orbital to virtual orbital as \cite{sahoo2004ab}
\begin{equation}
|\Psi_v\rangle=e^T\left\{1+S_v\right\}|\Phi_v\rangle, \label{eq1}
\end{equation}
where $T$ and $S_v$ can accounted for the singly and doubly excited-state configurations in our RCC theory (RCCSD method) ansatz as \cite{sahoo2004ab}
\begin{equation}
T=T_1+T_2 \qquad \text{and} \qquad S_v=S_{1v}+S_{2v} . \label{eq2}
\end{equation}

Accounting for excitations from both the core and valence orbitals of DF wave functions of Nb$^{4+}$ ion, we define the $T$ and $S_v$ operators 
in terms of the second quantization operators and their amplitudes $\rho$ as \cite{bijaya2005cc}
\begin{eqnarray}
T_{1}=\sum_{p, a} \rho_{pa} a^{\dagger}_p a_a , 
\  \
T_{2}=\frac{1}{4} \sum_{pq,ab}\rho_{pqab} a^{\dagger}_p a^{\dagger}_q a_b a_a , \nonumber \\
S_{1v}=\sum_{m\neq a} \rho_{p} a^{\dagger}_p a_v ,
\ \text{and} \
S_{2v}=\frac{1}{2} \sum_{pq,a}\rho_{pqva} a^{\dagger}_p a^{\dagger}_q a_a a_v , \ \
\end{eqnarray}
where the indices $p$ and $q$ range over a large number of virtual orbitals with energies up to 3500 a.u., and the indices $a$ and
$b$ range over all possible occupied orbitals. 

Finally, the matrix elements of a physical operator $\hat{O}$ between states $k$ and $v$ with the corresponding wave functions 
$|\Psi_{v}\rangle$ and $|\Psi_{k}\rangle$ are evaluated by \cite{nandy2014quadrupole,bijaya2005cc}
\begin{eqnarray}
O_{vk} &=& \frac{ \langle \Psi_v | \hat{O} | \Psi_k \rangle} {\sqrt{\langle \Psi_v | \Psi_v \rangle \langle \Psi_k | \Psi_k \rangle}} \nonumber \\
&=& \frac{\langle \Phi_v | \{S_v^{\dagger} +1 \} \overline{\hat{O}} \{ 1+ S_k \} |\Phi_k \rangle} {\langle \Phi_v | \{S_v^{\dagger} +1 \} \overline{\hat{N}} \{ 1+ S_k \} |\Phi_k \rangle},
\label{eq3}
\end{eqnarray}
where $\overline{\hat{O}}= e^{T^{\dagger}} \hat{O} e^{T}$  and $\overline{\hat{N}}= e^{T^\dagger} e^{T}$ with $\overline{\hat{O}}$ and 
$\overline{\hat{N}}$ being non-terminating series. In Eq. (\ref{eq3}), $\hat{O}$ can be replaced by the E1, M1 and electric quadrupole (E2) 
operators depending upon the matrix elements that need to be evaluated for the analysis.

{In this work, we have used Gaussian-type orbitals (GTOs) as the single-electron basis set in the calculations. The GTOs for the large ($L$) and small ($S$) components are given by \cite{boys1950electronic, wilson1997practical, mohanty1989kinetically}}
\begin{eqnarray}
g^{L}_{l,i}&=&N^Lr^{l+1}e^{-{\eta_i}r^2}\nonumber\\
\text{and }\nonumber\\
g^{S}_{l,i}&=&N^S\bigg[\frac{d}{dr}+\frac{\kappa}{r}\bigg]   g^{L}_{l,i}.  
\end{eqnarray}
{Here, $N^{L/S}$ represents the normalization constant for $L/S$ components, ${\eta_i}$ is an arbitrary coefficient suitably chosen for accurate calculations of wave functions, $l$ is the orbital quantum number and $\kappa$ is the relativistic quantum number. The exponents ${\eta_i}$s are determined by the relation ${\eta_i}={\eta_0}\beta^{i-1}$. We have used a large basis set of functions with 40, 39, 38, 37, 36, 35, and 34 GTOs for the $s, p, d, f, g, h$, and $i$ orbitals, respectively. We have given the list of ${\eta_0}$ and $\beta$ parameters that are used in present calculations for each symmetry in Table \ref{tab_basis}.}
\begin{table}[]
\centering
\caption{ List of ${\eta_0}$ and $\beta$ parameters used to define the GTOs for different symmetries to construct single-particle orbitals in the present calculations.}
\begin{tabular}{p{0.7cm}p{1cm}p{1cm}p{0.8cm}p{0.8cm}p{0.8cm}p{0.8cm}p{0.8cm}}
\hline
& $s$ & $p$ & $d$ & $f$& $g$ & $h$ & $i$ \\ [1ex]
\hline
${\eta_0}$ & 0.0009 & 0.0008 & 0.001 & 0.004 & 0.005 & 0.005 & 0.005 \\[1ex]
$\beta$  & 2.15 & 2.15 & 2.15 & 2.25 & 2.35 & 2.35 & 2.35 \\
\hline
\end{tabular}
\label{tab_basis}
\end{table}

\section{Estimation of Energies} \label{energies}
{In order to find out accuracy of the wave functions from the RCCSD method used to determine various matrix elements, we present energies obtained from this method in Table~\ref{taben}. Contributions from the Breit and QED corrections to the calculated energies are also shown explicitly in the table. This shows the Breit and QED contributions are within $0.01\%$ of the total calculated values. The final calculated values are also compared with those are tabulated in the NIST database \cite{ralchenko2008nist}. We find about $0.5\%$ variation between our calculated energies and NIST data.} 
\begin{table*}
\begin{center}
\caption{\label{taben}
Estimated energies (in cm$^{-1}$) of various states of Nb$^{4+}$ ion and their comparison with available NIST values. Percent deviation ($\delta(\%)$) is also provided. Values in square brackets depict the order of 10.}
\scalebox{1}[1]{
\begin{tabular}{ccccccccccc}
\hline\hline
Source & $4D_{3/2}$ & $4D_{5/2}$ & $4F_{5/2}$ & $4F_{7/2}$ & $5S_{1/2}$ & $5P_{1/2}$ & $5P_{3/2}$ & $6S_{1/2}$ & $6P_{1/2}$ & $6P_{3/2}$\\
\hline
RCC & 407832.95 & 405874.28 & 191705.65 & 191631.10 & 332251.04 & 278918.66 & 275246.78 & 179435.61 & 157407.86 & 155865.40 \\
$+$Breit & 37.10 & 118.95 & 53.42 & 58.32 & -67.19 & -112.15 & -56.71 & -27.85 & -47.92 & -24.78\\
$+$QED & 36.25 & 32.79 & 5.13 & 4.89 & -84.64 & 6.35 & 3.23 & -32.26 & 2.46 & 1.19\\
\hline
Total & 407906.30 & 406026.02 & 191764.20 & 191694.31 & 332099.21 & 278812.90 & 275193.30 & 179375.5 & 157362.40 & 155841.81\\
NIST  & 407897.00 & 406029.60 & 192637.90 & 192499.50 & 331967.40 & 278701.80 & 275097.00 & 179400.70 & 157390.50 & 155873.70\\
\hline
$\delta(\%)$ & 2.28[-3] & 8.86[-4] & 4.54[-1] & 4.18[-1] & 3.97[-2] & 3.99[-2] & 3.50[-2] & 1.40[-2] & 1.79[-2] & 2.05[-2]\\
\hline
\hline
\end{tabular}
}
\end{center}
\end{table*}

\section{Dipole Polarizabilities}\label{3}

Interaction of an atomic or ionic system with external electromagnetic fields leads to electric and magnetic polarization of system. This causes the 
shifts in energy levels of the considered atomic system, which can be estimated till the second-order by evaluating electric and magnetic dipole polarizabilities of the 
considered states. Thus, accurate determination of the E1 and M1 polarizabilities is essential in order to estimate possible systematics in the 
clock states of the considered atomic system. It requires determination of the E1 and M1 matrix elements and energies. The required matrix elements 
are evaluated using the aforementioned RCC method, whereas excitation energies are taken from the NIST database \cite{ralchenko2008nist} in this work. Further details of these calculations and obtained results are discussed below.

\subsection{Evaluation of E1 Polarizabilities} 

The total E1 polarizability ($\alpha^{E1}_{F_n}(\omega)$) of any hyperfine state $|F_n,M_F\rangle$ at any frequency ($\omega$), for linearly polarized
light for off-resonant spectrum can be represented in terms of scalar and tensor components as \cite{manakov1986atoms,singh2016comparing,schaffer2017towards,bonin1997dipole}
\begin{equation}
\label{eqdipoleF}
\alpha^{E1}_{F_n}(\omega)=\alpha_{n0,{F_n}}^{E1}(\omega)+\frac{3M_{F}^2-{F_n}({F_n}+1)}{{F_n}(2{F_n}-1)}\alpha_{n2,{F_n}}^{E1}(\omega) .
\end{equation}
Further, the scalar $\alpha_{n0,{F_n}}^{E1}(\omega)$ and tensor $\alpha_{n2,{F_n}}^{E1}(\omega)$ components are given in terms of total angular 
momentum ($J_n$) dependent polarizabilities by
\begin{eqnarray}\label{statF}
\alpha_{n0,{F_n}}^{E1}(\omega)&=&\alpha_{n0,{J_{n}}}^{E1}(\omega)
\end{eqnarray}
and
\begin{eqnarray}\label{eqc}
& &\alpha_{n2,{F_n}}^{E1}(\omega)=(-1)^{J_n+F_n+I} \left\{ \begin{array}{ccc}
 F_n & J_n & I\\
 J_n & F_n &2
 \end{array}\right\}\alpha_{n2,J_{n}}^{E1}(\omega) \nonumber \\
& \times & \sqrt{\frac{F_n(2F_n-1)(2F_n+1)(2J_n+3)(2J_n+1)(J_n+1)}{(2F_n+3)(F_n+1)(J_n)(2J_n-1)}}.
 \nonumber \\
 & &
\end{eqnarray}
Since calculating hyperfine level wave functions is challenging, the F-dependent dynamic polarizability for linearly polarized light can be 
determined using  Eq.~(\ref{eqdipoleF}) from the atomic E1 polarizabilities. In the above expressions, $\alpha_{n0,J_{n}}^{E1}(\omega)$ and 
$\alpha_{n2,J_{n}}^{E1}(\omega)$ can be determined using the relations
\begin{eqnarray}
\alpha_{n0,J_n}^{E1}(\omega)&=&-\frac{1}{3(2J_n+1)}\sum_{J_k}|\langle J_n||\textbf{D}||J_k \rangle|^2 \nonumber \\
 & \times & \left[\frac{1}{\delta E_{nk}+\omega}+\frac{1}{\delta E_{nk}-\omega}\right], \label{scalar}
\end{eqnarray}
and
\begin{eqnarray}
\alpha_{n2,J_n}^{E1}(\omega)&=&2\sqrt{\frac{5J_n(2J_n-1)}{6(J_n+1)(2J_n+3)(2J_n+1)}} \times \nonumber \\
& \times & \sum_{J_k}(-1)^{J_k+J_n+1}
 \left\{ \begin{array}{ccc}
 J_n& 2 & J_n\\
 1 & J_k &1
 \end{array}\right\}  \nonumber\\
 & \times & |\langle J_n||\textbf{D}||J_k \rangle|^2 \left[\frac{1}{\delta E_{nk}+\omega}+\frac{1}{\delta E_{nk}-\omega}\right]\label{tensor}. \ \ \ \
\end{eqnarray}
Here, $|\langle J_n||\textbf{D}|| J_k \rangle|$ are the reduced electric-dipole matrix elements (reduced using Wigner--Eckart theorem) with 
$\textbf{D}$ being the electric-dipole (E1) operator, $J_k$ being angular momentum of intermediate state $k$ and $\delta E_{nk}=E_n-E_k$ with 
$E_i^0$ correspond to the unperturbed energies of the states for $i=n,k$. The terms in curly bracket in Eqs.~(\ref{eqc}) and (\ref{tensor}) refer 
to 6-j symbols. The static values of the E1-polarizabilities can be deduced by substituting $\omega=0$ in the above expressions.



Above expressions for the E1-polarizabilities can be used to estimate contributions from the low-lying intermediate states whose matrix 
elements can explicitly be determined. To account for contributions from the core and continuum orbitals, we divide them into core 
($\alpha^c$), core-valence ($\alpha^{vc}$) and valence ($\alpha^{val}$) correlations as discussed in Ref. \cite{kaur2015properties}. Contributions 
from the virtual intermediate states correspond to $\alpha^{val}$, which are further divided into `Main' and `Tail' accounting 
contributions from the low-lying virtual states and high-lying virtual states, respectively. Contributions from the core orbitals without and 
with valence orbital interactions are given by $\alpha^c$ and $\alpha^{vc}$, respectively. Thus, we can write
\begin{equation} \label{vc}
{\alpha^{RCC}=\alpha^c+\alpha^{vc}+\alpha^{val},}
\end{equation}
where $\alpha^{val}=\alpha^{val}_{Main}+\alpha^{val}_{Tail}$.

{To verify the roles of the Breit and QED interactions in the determination of the E1 polarizabilities, we present the final value of $\alpha^{E1}$ as}
\begin{equation}
  {\alpha^{E1} =\alpha^{RCC+Breit+QED}.}
\end{equation}
{As mentioned above, energies are taken from the NIST data in the evaluation of the Main contributions. NES contributions to the length gauge expression for the E1 matrix elements are negligible~\cite{safronova2006multipole,johnson1995relativistic}, thus, contributions from the NESs to the final $\alpha^{E1}$ values are expected to be small.}

The results for J-dependent static E1 polarizabilities ($\omega=0$) of the $4D_{3/2}$ and $4D_{5/2}$ states from the off-resonant 
lasing spectra have been listed in Table~\ref{statpol}. From this, it can be observed that the $4D_{3/2}$--$5P_{1/2,3/2}$ and 
$4D_{3/2}$--$4F_{5/2}$ transitions contribute dominantly to the static E1 polarizabilities of the $4D_{3/2}$ state, whereas, 
the $4D_{5/2}$--$5P_{3/2}$ and $4D_{5/2}$--$4F_{5/2,7/2}$ transitions contribute majorly to the static E1 polarizability of the 
$4D_{5/2}$ state. As shown in Eq.~(\ref{statF}), the scalar component of the E1 polarizability for both the atomic and hyperfine 
levels are equal, thus, it can safely be deduced that the static scalar E1 polarizabilities $\alpha^{E1}_{n0,F_n}$ for the
$4D_{3/2}[|3,\pm 2\rangle]$ and $4D_{5/2}[|3,\pm 2\rangle]$ hyperfine levels are {$3.1429$ a.u. and $3.1506$ a.u., respectively. In
order to validate the accuracy of our results, we have compared our results with the static dipole polarizability values 
estimated using the RCC method without Breit and QED corrections as well as third-order relativistic many body perturbation theory (RMBPT3) method incorporating major physical 
correlations specifically through random-phase approximation, Br\"uckner orbitals, structural radiations as well as 
normalization of wave functions. The comparison of results showed an underlying uncertainties $\delta_1(\%)$ and $\delta_2(\%)$, less than $0.5\%$ and $10\%$ with respect to RCC and RMPBT3 methods, respectively. The high uncertainty in comparison with RMBPT3 method is expected due to the fact that this method incorporates only major physical correlations upto third-order only. Besides, when we include the near-resonant lasing spectra, it is observed that the linewidths of the contributing states to the E1-polarizabilities are negligibly small (of the order of $10^{-6}$ or less).}

\begin{table*}[t]
\caption{Contributions from different dominant far-off resonant transitions to the static dipole polarizabilities (in a.u.) of the $4D_{3/2}$ and $4D_{5/2}$ states of Nb$^{4+}$ ion. Percent deviations $(\delta_1 (\%))$ and $(\delta_2 (\%))$ represent the deviation in RCC+Breit+QED results with RCC and RMBPT3 results, respectively. {Here values under `d' denote the E1 matrix elements. $\alpha_{w0}(0)$ and $\alpha_{w2}(0)$ correspond to the scalar and tensor components, respectively, of the E1 polarizability for the $4D_{3/2}$ state, while $\alpha_{v0}(0)$ and $\alpha_{v2}(0)$ denote the same for the $4D_{5/2}$ state.}}
\scalebox{1}[1]{
\begin{tabular}{cccccccc}
\hline
\hline
\multicolumn{4}{c}{$4D_{3/2}$} & \multicolumn{4}{c}{$4D_{5/2}$}\\
Transition & d & {$\alpha_{w0}(0)$} & {$\alpha_{w2}(0)$} & Transition & d & {$\alpha_{v0}(0)$} & {$\alpha_{v2}(0)$}\\
\hline
$4D_{3/2}$--$5P_{1/2}$&1.179&0.3936&-0.3936 & $4D_{5/2}$--$5P_{3/2}$&1.569&0.4585&-0.4585\\
$4D_{3/2}$--$6P_{1/2}$&0.227&0.0075&-0.0075 & $4D_{5/2}$--$6P_{3/2}$&0.319&0.0099&-0.0099\\
$4D_{3/2}$--$7P_{1/2}$&0.096&0.0011&-0.0011 & $4D_{5/2}$--$7P_{3/2}$&0.158&0.0020&-0.0020\\
$4D_{3/2}$--$8P_{1/2}$&0.065&0.0005&-0.0005 & $4D_{5/2}$--$8P_{3/2}$&0.095&0.0007&-0.0007\\
$4D_{3/2}$--$9P_{1/2}$&0.034&0.0001&-0.0001 & $4D_{5/2}$--$9P_{3/2}$&0.063&0.0003&0.0003\\
$4D_{3/2}$--$10P_{1/2}$&0.028&0.0001&-0.0001 & $4D_{5/2}$--$10P_{3/2}$&0.045&0.0001&0.0001\\
$4D_{3/2}$--$5P_{3/2}$&0.515&0.0730&0.0584 & $4D_{5/2}$--$4F_{5/2}$&0.480&0.0263&0.0301\\
$4D_{3/2}$--$6P_{3/2}$&0.106&0.0016&0.0013 & $4D_{5/2}$--$5F_{5/2}$&0.064&0.0004&0.0005\\
$4D_{3/2}$--$7P_{3/2}$&0.053&0.0003&0.0002 & $4D_{5/2}$--$6F_{5/2}$&0.043&0.0001&0.0001\\
$4D_{3/2}$--$8P_{3/2}$&0.032&0.0001&0.0001 & $4D_{5/2}$--$4F_{7/2}$&2.158&0.5318&-0.1899\\
$4D_{3/2}$--$9P_{3/2}$&0.021&~0.0&~0.0 & $4D_{5/2}$--$5F_{7/2}$&0.303&0.0081&-0.0029\\
$4D_{3/2}$--$10P_{3/2}$&0.015&~0.0&~0.0 & $4D_{5/2}$--$6F_{7/2}$&0.167&0.0021&-0.0008\\
$4D_{3/2}$--$4F_{5/2}$&1.770&0.5324&-0.1065&&&&\\
$4D_{3/2}$--$5F_{5/2}$&0.237&0.0074&-0.0015&&&&\\
$4D_{3/2}$--$6F_{5/2}$&0.157&0.0028&-0.0006&&&&\\
$4D_{3/2}$--$7F_{5/2}$&0.311&0.0105&-0.0021&&&&\\
$4D_{3/2}$--$8F_{5/2}$&0.265&0.0073&-0.0015&&&&\\
\hline
$\alpha^{val}_{Main}$ &&{1.0383}&-0.4551& $\alpha^{val}_{Main}$ && {1.0403}&{-0.6336}\\
$\alpha^{val}_{Tail}$&&0.0393&-0.0110 & $\alpha^{val}_{Tail}$ && 0.0603&-0.0216 \\
$\alpha^{vc}$ &&-0.2148&0.1450 & $\alpha^{vc}$ && -0.2297&0.2297\\
$\alpha^c$ &&2.2780 & & $\alpha^{c}$ && 2.2780&\\
\hline
RCC &&{3.1408}&-0.3211& RCC && {3.1489}& {-0.4255}\\[0.5ex]
RCC+Breit+QED ($\alpha^{E1}$) && 3.1429  & -0.3201 & RCC+Breit+QED ($\alpha^{E1}$) && 3.1506 & -0.4251 \\[0.5ex]
RMBPT3 &&3.3540& -0.3406 & RMBPT3 && 3.3766 & -0.4105\\
\hline
$\delta_1 (\%)$ && 0.07 & 0.31 &$\delta_2 (\%)$ && 0.05 & 0.09\\
$\delta_2 (\%)$ && {6.79} & 6.40 & $\delta_1 (\%)$ && {7.17} & {3.43}\\
\hline
\hline
\end{tabular}
}
\label{statpol}
\end{table*}

\subsection{Evaluation of M1 Polarizabilities}
{The interaction of magnetic dipoles $\mu_m$ within an ionic system with an external magnetic field $\mathcal{B}$ leads to the magnetic polarization of the system. These magnetic interactions can be addressed quantum mechanically, using the M1 operator $\hat{O}^{M1}=(\textbf{L}+2\textbf{S})\mu_B$ for Russel-Saunders coupling, with $\mu_B$ being the Bohr magneton and $\textbf{L}$ and $\textbf{S}$ being the orbital and spin angular momentum operators, respectively. Quantitatively, the phenomenon of magnetic polarization can be described through the physical quantity $\alpha^{M1}_{n,F_n}$, generally known as magnetic dipole polarizability, which, for any hyperfine state $n$, can be expressed as}
\cite{pan2020optical}
\begin{equation}
\label{alpham1}
\alpha^{M1}_{n,F_n}=-\frac{2}{3(2F_n+1)}\sum_{k}\frac{|\langle F_n||\hat{O}^{M1}||F_k\rangle|^2}{E_{F_n}-E_{F_{k}}},
\end{equation} 
where $\hat{O}^{M1}$ is the M1 operator and $F_k$ represents the intermediate hyperfine states to which all the allowed transitions from $F_n$ are 
possible. In Eq.~(\ref{alpham1}), energies of the hyperfine levels can be estimated by~\cite{sahoo2015correlation}
\begin{eqnarray}\label{energyhyp}
E_{F_n}&=&\frac{A_{{\rm{hf}}}}{2}K_n \nonumber \\
  && + B_{{\rm{hf}}}\left[\frac{\frac{3}{4}K_n(K_n+1)-I(I+1)J_n(J_n+1)}{2I(2I-1)J_n(2J_n-1)}\right],\nonumber\\  
\end{eqnarray}
where $K_n=F_n(F_n+1)-J_n(J_n+1)-I(I+1)$, and $A_{{\rm{hf}}}$ and $B_{{\rm{hf}}}$ are the M1 and E2 hyperfine interaction constants for the 
considered states, respectively, which are given by~\cite{sahoo2015correlation}
\begin{eqnarray}\label{eqhf1}
A_{{\rm{hf}}}=\mu_N g_I \frac{\langle J_n||\hat{O}_{hf}^{(1)}||J_n\rangle}{\sqrt{J_n(J_n+1)(2J_n+1)}} \nonumber\\
\end{eqnarray}
and
\begin{eqnarray}\label{eqhf2}
B_{{\rm{hf}}}&=& 2Q\left[\frac{2J_n(2J_n-1)}{(2J_n+1)(2J_n+2)(2J_n+3)}\right]^{\frac{1}{2}}\langle J_n||\hat{O}_{hf}^{(2)}||J_n\rangle,\nonumber\\ &&
\end{eqnarray}
with $\mu_N$ being the nuclear Bohr magneton; $g_I=\frac{\mu_I}{I}$, with nuclear M1 moment $\mu_I$ and and nuclear spin $I$, and $Q$ is the nuclear 
E2 moment. {In Eqs.~(\ref{eqhf1}) and~(\ref{eqhf2}), $\hat{O}_{hf}^{(1)}$ and $\hat{O}_{hf}^{(2)}$ depict hyperfine operators given in the form~\cite{schwartz1955hfs}}
\begin{eqnarray}
    \hat{O}_{hf}^{(1)}=\sum_{j} -ie\sqrt{\frac{8\pi}{3}}\frac{1}{r_j^2}\alpha_j\rm{Y}^{(0)}_{1q}(r_j)
\end{eqnarray}
 and
 \begin{eqnarray}
     \hat{O}_{hf}^{(2)}=\sum_{j} -ie\frac{1}{r_j^3}\rm{C}_q^{(2)}(r_j),
 \end{eqnarray}
{where, $q$ represents the order of spherical harmonics and functions $\rm Y$ and $\rm C$ are the tensor operators with respective orders given in the parentheses as superscripts.}
Further, the reduced F-dependent M1-matrix elements, (reduced using Wigner-Eckart theorem) in Eq.~(\ref{alpham1}), can easily be determined 
by using J-dependent M1 matrix {elements as~\cite{sobelman1979angular,lindgren2012atomic}}
\begin{eqnarray}
\langle F_n||\hat{O}^{M1}||F_k\rangle=\sqrt{(2F_n+1)(2F_k+1)}\times\nonumber\\
\left\lbrace\begin{array}{ccc}
J_n & I & F_n\\
F_k & 1 & J_n
\end{array}\right\rbrace \langle J_n||\hat{O}^{M1}||J_k\rangle .
\end{eqnarray}

{Contributions from the Breit and QED effects to the calculated M1 matrix elements are found to be extremely small. The uncertainties to the M1 polarizabilities mostly rely on the estimated energy differences which are determined from the calculated hyperfine structure constants using the RCCSD method. These uncertainties will dominate over the contributions from the NESs. Again, it has been shown that contributions from NESs to the evaluation of the M1 matrix elements drastically decrease with increasing atomic number~\cite{johnson1995relativistic,indelicato1995neg}. Therefore, NES contributions to the evaluation of M1 polarizabilities can be safely neglected here.}

On the basis of selection rules of the M1 transition, it is found that the $4D_{3/2}|F=3\rangle$--$4D_{3/2}|F=4\rangle$ transition 
is the primary contributor to the M1 polarizability of the $4D_{3/2}|F=3\rangle$ state, with an estimated value of {$2.6836 \times 10^{-23}$ 
JT$^{-2}$. Similarly, for the $4D_{5/2}|F=3\rangle$ state, the $4D_{5/2}|F=3\rangle$--$4D_{5/2}|F=2,4\rangle$ transitions contribute to its 
M1 polarizability, yielding a value of $9.0125 \times 10^{-23}$ JT$^{-2}$. Eventually, the validation of our results is based on the comparison 
of our results to the $\alpha^{M1}$ values obtained using RCC and RMBPT3 methods resulting in the corresponding percent deviations of $1.86\%$ and $4.55\%$ and $3.65\%$ and $4.59\%$ for 
the $4D_{3/2}$ and $4D_{5/2}$ states, respectively.}

\section{Estimation of Major Systematic Shifts to Clock States}\label{4}


{We would like now to estimate and discuss major systematics that contribute mostly to the uncertainty of the clock frequency measurement. We use the calculated quantities discussed in the previous section to estimate the BBR, Zeeman, electric quadrupole and Doppler shifts. All these estimations are summarized in Table~\ref{tabconc} and are discussed below.} 

\subsection{BBR Shifts}

The impact of temperature $T$ of the environment to the clock frequency measurement is prevalent and leads to the thermal fluctuations of 
electromagnetic field. These thermal fluctuations are further experienced by an atomic system resulting in the interaction of system with both 
electric and magnetic field components of blackbody radiations, thus inducing the shifts in the energy states, generally known as BBR Stark (BBRS) and 
BBR Zeeman (BBRZ) shifts, respectively. They are one of the major irreducible contributions to uncertainty of any atomic clock~\cite{farley1981accurate,yu2018selected}. 

Considering system at the room temperature T$=300$ K, the {BBRS} shift of an energy level can be expressed in terms of differential static scalar
polarizability $\Delta\alpha_{0}^{E1}=\alpha_{v0,F_v}^{E1}-\alpha_{w0,F_w}^{E1}$, of the considered clock transition as~\cite{arora2007blackbody}
\begin{eqnarray}\label{eqbbrstark}
\Delta\nu_{\rm BBR}^{\rm E1}=-\frac{1}{2h}(831.9\rm{V/m})^2 \Delta\alpha_0^{E1} ,
\end{eqnarray}
where, in Eq.~(\ref{eqbbrstark}), the E1 polarizability {$\alpha_{n0,F_n}^{E1}$} in a.u. can be converted into SI units using
$\alpha/h(Hz(V/m)^{-2})=2.48832\times10^{-8}\alpha(a.u.)$.

Similarly, the {BBRZ} shift due to the differential M1 polarizability $\Delta \alpha^{M1}$ of the clock transition is given by~\cite{arora2012multipolar}
\begin{equation}\label{eqbbrmag}
\Delta\nu_{\rm BBR}^{\rm M1}=-\frac{1}{2h}(2.77\times 10^{-6} \rm{T})^2 \Delta\alpha^{M1},
\end{equation}
at $300$ K. Here, $\Delta \alpha^{M1}=\alpha^{M1}_{v,F_v}-\alpha^{M1}_{w,F_w}$ (in $\mu_B$) can be converted into SI units using the relation $1\mu_B=9.274\times 10^{-24}$JT$^{-1}$.

From Table~\ref{statpol}, it can be seen that the scalar static E1 polarizabilities $\alpha^{E1}_{n0,F_n}$ for the $4D_{3/2}[|3,\pm 2\rangle]$ and 
$4D_{5/2}[|3,\pm 2\rangle]$ hyperfine levels are about {$3.1429$ a.u. and $3.1506$ a.u.}, respectively. Thus, the differential static E1 polarizability 
is estimated to be {$7.7000\times 10^{-3}$ a.u.,} resulting in {BBRS} and fractional {BBRS} shifts of {$-6.6299\times 10^{-5}$ Hz and 
$-1.1834\times10^{-18}$, respectively, with an uncertainty of $5.19\%$ and $5.20\%$ w.r.t. to BBRS and fractional BBRS shift evaluated using RCC method only}. However, for the {BBRZ} shift, the evaluated M1-polarizabilities for the $4D_{3/2}|F=3\rangle$ and 
$4D_{5/2}|F=3\rangle$ hyperfine levels are found to be {$2.6836\times 10^{-23}$ JT$^{-2}$ and $9.0125\times 10^{-23}$ JT$^{-2}$, respectively with the corresponding uncertainties of $1.86\%$ and $4.55\%$ with respect to the M1 polarizability values evaluated using RCC approach only.
This corresponds to the BBRZ shift ($\Delta\nu^{M1}_{BBR}$) of $-0.3664$ Hz with an uncertainty of $-0.0029$ Hz, whereas, the fractional BBRZ shift 
is found to be $-6.5402\times 10^{-15}$ resulting in an estimated uncertainty of $-5.1765\times 10^{-17}$ w.r.t. RCC results.
}
\subsection{Zeeman Shifts}

The interaction of the M1 moment $\mu_B$ of an ion with an external magnetic beam of intensity $\mathcal{B}$ arises significant shifts in the
hyperfine energy levels as well as transition frequencies of the system; usually known as Zeeman shifts~\cite{campbell2012single}. The linear Zeeman 
shifts can be avoided if average is taken over the transition frequencies with positive and negative $M_F$ states, as described in 
Refs.~\cite{dzuba2021time,takamoto2006improved}. Although the first-order Zeeman shifts are avoidable, the quadratic Zeeman shifts can contribute 
significantly to the frequency uncertainty budget and hence, they must be estimated. The quadratic Zeeman shift to the clock transition can be 
expressed in terms of differential M1 polarizability $\Delta\alpha^{M1}$ as~\cite{porsev2020calculation}
\begin{equation}
\label{eqnuz2}
\Delta\nu^{(Z2)}=-\frac{1}{2h}\Delta\alpha^{M1}\mathcal{B}^2 
\end{equation}
with $\Delta\alpha^{M1}=\alpha^{M1}_{v,F_v}-\alpha^{M1}_{w,F_w}$. Substituting the M1 polarizability values of $2.6336\times 10^{-23}$ JT$^{-2}$ and 
$9.4223\times 10^{-23}$ JT$^{-2}$ corresponding to the $4D_{3/2}|F=3\rangle$ and $4D_{5/2}|F=3\rangle$ hyperfine levels lead to the quadratic Zeeman 
shift ({$\Delta\nu^{(Z2)}$}) and fractional Zeeman shift of the clock transition as $-5.1197\times10^{-6}$ Hz and $-9.1387\times 10^{-20}$, respectively.

\begin{table*}[t]
\begin{center}
\caption{\label{tabconc}%
Estimated systematic shifts in the $4D_{3/2}|3,\pm2\rangle$--$4D_{5/2}|3,\pm2\rangle$ THz clock transition of the Nb$^{4+}$ ion.}
\scalebox{1}[1]{
\begin{tabular}{ccc}
\hline
Source & $\Delta\nu$ (Hz) & $\frac{\Delta\nu}{\nu_0}$\\
\hline
Electric Quadrupole & $0$ & $0$\\
BBR Zeeman (T=300 K) & $-0.3664 $ & $-6.5402\times 10^{-15}$\\
BBR Stark (T=300 K) & $-6.6299\times 10^{-5}$ & $-1.1834\times 10^{-18}$\\
Quadratic Zeeman (B=$10^{-8}$ T)& $-5.1197\times 10^{-6}$ & $-9.1387\times 10^{-20}$\\
Second-order Doppler (Thermal) & $-2.5081\times 10^{-14}$ & $-4.4769\times 10^{-28}$\\
\hline
\end{tabular}
}
\end{center}
\end{table*}

\subsection{Electric Quadrupole Shifts}

The interaction of quadrupole moments of the clock levels with the residual electric-field gradient at the trap center can be described 
using the Hamiltonian $H_Q$~\cite{itano2000external}, given by
\begin{eqnarray}
H_Q=\nabla\textbf{E}^{(2)}.\textbf{$\Theta$}^{(2)},
\end{eqnarray}
where $\nabla$\textbf{E}$^{(2)}$ is the tensor describing the gradient of external electric field and \textbf{$\Theta$}$^{(2)}$ is the electric-quadrupole operator for the considered ion. This interaction further leads to the shift in the energies of the clock levels, that can be estimated using~\cite{porsev2020optical}
\begin{eqnarray} \label{eqQ}
\Delta E\simeq -\frac{1}{2}\Delta\langle Q_0 \rangle \frac{\partial\mathcal{E}_z}{\partial z}.
\end{eqnarray}
In Eq.~(\ref{eqQ}), $\langle Q_0 \rangle\equiv\langle \gamma JIFM|{\textbf{$\Theta$}^{(2)}}|\gamma JIFM\rangle$, with $\gamma$ including all other quantum 
numbers. The expectation value $\langle Q_0 \rangle$ of any state $n$ can be evaluated using~\cite{itano2000external}
\begin{eqnarray}\label{exp}
\langle\gamma JIFM|Q_0|\gamma JIFM\rangle=(-1)^{I+J_n+F_n}[3M_F^{2}-F_n(F_n+1)]\nonumber\\
\times\sqrt{\frac{2F_n+1}{(2F_n+3)(F_n+1)(2F_n-1)}}\nonumber\\
\times\left\{\begin{array}{ccc}
J_n & J_n & 2\\
F_n & F_n & I
\end{array}\right\}\times\langle\gamma J||\textbf{$\Theta$}^{(2)}||\gamma J\rangle,\nonumber\\
\end{eqnarray}
with $\langle\gamma J||\Theta^{(2)}||\gamma J\rangle$ being the reduced matrix element of quadrupole operator. From Eq.~(\ref{exp}), it 
can easily be deduced that minimization of electric quadrupole shift can be observed if $3M_F^2-F_n(F_n+1)=0$, implying
$3M_F^2=F_n(F_n+1)$. Since Nb$^{4+}$ has nuclear spin, $I=\frac{9}{2}$, thus total angular momentum $F$ acquires values  $3$ to $6$ and
$2$ to $7$ for the $4D_{3/2}$ and $4D_{5/2}$ states, respectively. Therefore, the electric quadrupole shifts for both of these clock levels 
can nullify when $|F,M_F\rangle$ are chosen to be $|3,\pm 2\rangle$. This is the reason why we have selected the  $|3,\pm 2\rangle$ hyperfine 
levels in scheme, shown in Fig.~\ref{figclock}, for the clock frequency measurement. 

\subsection{Doppler Shift}

Doppler shifts occur due to distribution of velocity among the considered atomic systems in the experiment when an atomic clock system is operated 
on a particular temperature~\cite{sahoo2015springer}. The cold but moving ions in this case can interact with the field inside the microwave cavity 
with a spatial phase variation, thereby shifting the energy levels involved in the clock transition~\cite{guena2011doppler}. The accuracy of the 
atomic clock is limited by these Doppler shifts, hence, it is necessary to minimize both the first- and second-order Doppler shifts for 
high-precision clock frequency measurement. The first-order Doppler shift can be eliminated by using two probe beams in opposite directions during
the detection~\cite{wineland2013nobel}, however the second-order Doppler shift due to secular motion is quite considerable and can be as large
as $1$ part in $10^8$~\cite{guena2011doppler}. This second-order Doppler shift can be expressed in terms of mass $m$ of the considered ion and speed 
of light $c$ in vacuum as~\cite{zhang2017direct}
\begin{equation} \label{eqdoppler}
\Delta\nu_{\rm{D2}}=-\left(\frac{3\hbar\Gamma}{4mc^2}\right)\nu_0.
\end{equation}
Considering the current advanced experimental techniques, we recommend using cooling lasers under optimized working conditions for cooling the ion trap to the temperature closer to the Doppler-cooling limit ($T_D$). Acquiring temperature near $T_D$, the second-order Doppler shift due to the secular motion of the ion can further be minimized~\cite{huang2022liquid}. This Doppler-cooling limit depends on the natural linewidth of the atomic transition ($\Gamma^{-1}$) and is determined using the formula~\cite{phillips1998laser}
\begin{equation}\label{coollt}
T_D=\frac{\hbar\Gamma}{2K_B}.
\end{equation}
Here, $\Gamma$ is the rate of spontaneous emission of the excited state and is given by $\Gamma=\frac{1}{\tau_v}$, where $v$ corresponds to the excited state.
Substituting the value of Doppler cooling limit from Eq.~(\ref{coollt}) in Eq.~(\ref{eqdoppler}), we get
\begin{equation}\label{eqdop}
\Delta\nu_{\rm{D2}}=-\left(\frac{3K_B T_D}{2mc^2}\right)\nu_0.
\end{equation} 
For Nb$^{4+}$, the lifetime of excited state, i.e., $\tau_{4D_{5/2}}=12.65$ s, resulting to the Doppler cooling limit of $0.302$ pK. Consequently, 
substituting the value of $T_D$ in Eq.~(\ref{eqdop}), the second-order Doppler shift and fractional frequency shift are found to be 
$-2.5081 \times 10^{-14}$ Hz and $-4.4769 \times 10^{-28}$, respectively. 

{It is obvious from the above results that the BBRZ shift contributes at the $10^{-15}$ level while other systematic effects can contribute below $10^{-17}$. The uncertainty due to the BBRZ shift can be reduced further by performing the experiment at lower temperatures or measuring the M1 polarizability more precisely. Nonetheless, the large value of the BBRZ shift can be useful. Since it is highly sensitive to magnetic field fluctuations, the Nb$^{4+}$ ion clock can be utilized for quantum thermometry.}

\section{Conclusion}\label{5}

This work demonstrates that the $4D_{3/2}|F=3,M_F=\pm2\rangle\rightarrow 4D_{5/2},|F=3,M_F=\pm2\rangle$ transition in $^{93}$Nb$^{4+}$ ion can 
work as a prospective terahertz atomic clock. In view of this, we have explicitly discussed our clock proposal and evaluated major systematics 
to this transition including BBR, electric quadrupole, second-order Doppler as well as second-order Zeeman shifts. Maximum contribution in the 
systematics of this transition is observed to be contributed by external magnetic effects leading to considerable BBR Zeeman and quadratic Zeeman 
shifts. The minimization of other systematic shifts can be carried out by optimizing the parameters as suggested in the study. We also conclude that Nb$^{4+}$ THz clock is more sensitive to external magnetic fields in comparison to our previously proposed Zr$^{3+}$ THz clock. {Moreover, this clock scheme is independent of AC Stark shift and electric quadrupole shift, which makes its implementation more favourable for experimentalists. Therefore, we expect it to be more useful for evolving quantum thermometry techniques as well as for evolving terahertz devices including sources, detectors and THz spectrometers upon the successful implementation and development in an optimized environment.}

\section*{Acknowledgement}

This research was funded by the Innovation Program for Quantum Science and Technology (2021ZD0303200) and the National Natural Science Foundation
of China (Grant No. 624B2010). Research at Perimeter Institute is supported in part by the Government of Canada through the Department of Innovation, Science and Economic 
Development and by the Province of Ontario through the Ministry of Colleges and Universities. We acknowledge ParamVikram 1000 HPC facility at 
Physical Research Laboratory, Ahmedabad, India for carrying out relativistic coupled-cluster calculations. AC and BKS were supported by the 
Department of Space, Government of India to carry out this work.

\bibliography{NbV.bib}

\begin{thebibliography}{81}
\expandafter\ifx\csname natexlab\endcsname\relax\def\natexlab#1{#1}\fi
\expandafter\ifx\csname bibnamefont\endcsname\relax
  \def\bibnamefont#1{#1}\fi
\expandafter\ifx\csname bibfnamefont\endcsname\relax
  \def\bibfnamefont#1{#1}\fi
\expandafter\ifx\csname citenamefont\endcsname\relax
  \def\citenamefont#1{#1}\fi
\expandafter\ifx\csname url\endcsname\relax
  \def\url#1{\texttt{#1}}\fi
\expandafter\ifx\csname urlprefix\endcsname\relax\def\urlprefix{URL }\fi
\providecommand{\bibinfo}[2]{#2}
\providecommand{\eprint}[2][]{\url{#2}}

\bibitem[{\citenamefont{Derevianko and Pospelov}(2014)}]{derevianko2014hunting}
\bibinfo{author}{\bibfnamefont{A.}~\bibnamefont{Derevianko}} \bibnamefont{and} \bibinfo{author}{\bibfnamefont{M.}~\bibnamefont{Pospelov}}, \bibinfo{journal}{Nature Physics} \textbf{\bibinfo{volume}{10}}, \bibinfo{pages}{933} (\bibinfo{year}{2014}), \urlprefix\url{https://www.nature.com/articles/nphys3137}.

\bibitem[{\citenamefont{Rosenband et~al.}(2008)\citenamefont{Rosenband, Hume, Schmidt, Chou, Brusch, Lorini, Oskay, Drullinger, Fortier, Stalnaker et~al.}}]{rosenband2008frequency}
\bibinfo{author}{\bibfnamefont{T.}~\bibnamefont{Rosenband}}, \bibinfo{author}{\bibfnamefont{D.}~\bibnamefont{Hume}}, \bibinfo{author}{\bibfnamefont{P.}~\bibnamefont{Schmidt}}, \bibinfo{author}{\bibfnamefont{C.-W.} \bibnamefont{Chou}}, \bibinfo{author}{\bibfnamefont{A.}~\bibnamefont{Brusch}}, \bibinfo{author}{\bibfnamefont{L.}~\bibnamefont{Lorini}}, \bibinfo{author}{\bibfnamefont{W.}~\bibnamefont{Oskay}}, \bibinfo{author}{\bibfnamefont{R.~E.} \bibnamefont{Drullinger}}, \bibinfo{author}{\bibfnamefont{T.~M.} \bibnamefont{Fortier}}, \bibinfo{author}{\bibfnamefont{J.~E.} \bibnamefont{Stalnaker}}, \bibnamefont{et~al.}, \bibinfo{journal}{Science} \textbf{\bibinfo{volume}{319}}, \bibinfo{pages}{1808} (\bibinfo{year}{2008}).

\bibitem[{\citenamefont{Mehlstäubler et~al.}(2018)\citenamefont{Mehlstäubler, Grosche, Lisdat, Schmidt, and Denker}}]{Mehlstaubler_2018}
\bibinfo{author}{\bibfnamefont{T.~E.} \bibnamefont{Mehlstäubler}}, \bibinfo{author}{\bibfnamefont{G.}~\bibnamefont{Grosche}}, \bibinfo{author}{\bibfnamefont{C.}~\bibnamefont{Lisdat}}, \bibinfo{author}{\bibfnamefont{P.~O.} \bibnamefont{Schmidt}}, \bibnamefont{and} \bibinfo{author}{\bibfnamefont{H.}~\bibnamefont{Denker}}, \bibinfo{journal}{Reports on Progress in Physics} \textbf{\bibinfo{volume}{81}}, \bibinfo{pages}{064401} (\bibinfo{year}{2018}), \urlprefix\url{https://dx.doi.org/10.1088/1361-6633/aab409}.

\bibitem[{\citenamefont{McGrew et~al.}(2018)\citenamefont{McGrew, Zhang, Fasano, Sch{\"a}ffer, Beloy, Nicolodi, Brown, Hinkley, Milani, Schioppo et~al.}}]{mcgrew2018atomic}
\bibinfo{author}{\bibfnamefont{W.}~\bibnamefont{McGrew}}, \bibinfo{author}{\bibfnamefont{X.}~\bibnamefont{Zhang}}, \bibinfo{author}{\bibfnamefont{R.}~\bibnamefont{Fasano}}, \bibinfo{author}{\bibfnamefont{S.}~\bibnamefont{Sch{\"a}ffer}}, \bibinfo{author}{\bibfnamefont{K.}~\bibnamefont{Beloy}}, \bibinfo{author}{\bibfnamefont{D.}~\bibnamefont{Nicolodi}}, \bibinfo{author}{\bibfnamefont{R.}~\bibnamefont{Brown}}, \bibinfo{author}{\bibfnamefont{N.}~\bibnamefont{Hinkley}}, \bibinfo{author}{\bibfnamefont{G.}~\bibnamefont{Milani}}, \bibinfo{author}{\bibfnamefont{M.}~\bibnamefont{Schioppo}}, \bibnamefont{et~al.}, \bibinfo{journal}{Nature} \textbf{\bibinfo{volume}{564}}, \bibinfo{pages}{87} (\bibinfo{year}{2018}).

\bibitem[{\citenamefont{Kolkowitz et~al.}(2016)\citenamefont{Kolkowitz, Pikovski, Langellier, Lukin, Walsworth, and Ye}}]{kolkowitz2016gw}
\bibinfo{author}{\bibfnamefont{S.}~\bibnamefont{Kolkowitz}}, \bibinfo{author}{\bibfnamefont{I.}~\bibnamefont{Pikovski}}, \bibinfo{author}{\bibfnamefont{N.}~\bibnamefont{Langellier}}, \bibinfo{author}{\bibfnamefont{M.~D.} \bibnamefont{Lukin}}, \bibinfo{author}{\bibfnamefont{R.~L.} \bibnamefont{Walsworth}}, \bibnamefont{and} \bibinfo{author}{\bibfnamefont{J.}~\bibnamefont{Ye}}, \bibinfo{journal}{Phys. Rev. D} \textbf{\bibinfo{volume}{94}}, \bibinfo{pages}{124043} (\bibinfo{year}{2016}), \urlprefix\url{https://link.aps.org/doi/10.1103/PhysRevD.94.124043}.

\bibitem[{\citenamefont{Loeb and Maoz}(2015)}]{loeb2015using}
\bibinfo{author}{\bibfnamefont{A.}~\bibnamefont{Loeb}} \bibnamefont{and} \bibinfo{author}{\bibfnamefont{D.}~\bibnamefont{Maoz}}, \bibinfo{journal}{arXiv preprint arXiv:1501.00996}  (\bibinfo{year}{2015}).

\bibitem[{\citenamefont{Graham et~al.}(2013)\citenamefont{Graham, Hogan, Kasevich, and Rajendran}}]{graham2013new}
\bibinfo{author}{\bibfnamefont{P.~W.} \bibnamefont{Graham}}, \bibinfo{author}{\bibfnamefont{J.~M.} \bibnamefont{Hogan}}, \bibinfo{author}{\bibfnamefont{M.~A.} \bibnamefont{Kasevich}}, \bibnamefont{and} \bibinfo{author}{\bibfnamefont{S.}~\bibnamefont{Rajendran}}, \bibinfo{journal}{Phys. Rev. Lett.} \textbf{\bibinfo{volume}{110}}, \bibinfo{pages}{171102} (\bibinfo{year}{2013}).

\bibitem[{\citenamefont{Dzuba et~al.}(2018)\citenamefont{Dzuba, Flambaum, and Schiller}}]{dzuba2018testing}
\bibinfo{author}{\bibfnamefont{V.}~\bibnamefont{Dzuba}}, \bibinfo{author}{\bibfnamefont{V.}~\bibnamefont{Flambaum}}, \bibnamefont{and} \bibinfo{author}{\bibfnamefont{S.}~\bibnamefont{Schiller}}, \bibinfo{journal}{Phys. Rev. A} \textbf{\bibinfo{volume}{98}}, \bibinfo{pages}{022501} (\bibinfo{year}{2018}).

\bibitem[{\citenamefont{Tonouchi}(2007)}]{tonouchi2007cutting}
\bibinfo{author}{\bibfnamefont{M.}~\bibnamefont{Tonouchi}}, \bibinfo{journal}{Nature photonics} \textbf{\bibinfo{volume}{1}}, \bibinfo{pages}{97} (\bibinfo{year}{2007}).

\bibitem[{\citenamefont{Kulesa}(2011)}]{kulesa2011terahertz}
\bibinfo{author}{\bibfnamefont{C.}~\bibnamefont{Kulesa}}, \bibinfo{journal}{IEEE Transactions on Terahertz Science and Technology} \textbf{\bibinfo{volume}{1}}, \bibinfo{pages}{232} (\bibinfo{year}{2011}).

\bibitem[{\citenamefont{Norrgard et~al.}(2021)\citenamefont{Norrgard, Eckel, Holloway, and Shirley}}]{norrgard2021quantum}
\bibinfo{author}{\bibfnamefont{E.~B.} \bibnamefont{Norrgard}}, \bibinfo{author}{\bibfnamefont{S.~P.} \bibnamefont{Eckel}}, \bibinfo{author}{\bibfnamefont{C.~L.} \bibnamefont{Holloway}}, \bibnamefont{and} \bibinfo{author}{\bibfnamefont{E.~L.} \bibnamefont{Shirley}}, \bibinfo{journal}{New Journal of Physics} \textbf{\bibinfo{volume}{23}}, \bibinfo{pages}{033037} (\bibinfo{year}{2021}).

\bibitem[{\citenamefont{Kim et~al.}(2019)\citenamefont{Kim, Wang, Hu, and Han}}]{kim2019chip}
\bibinfo{author}{\bibfnamefont{M.}~\bibnamefont{Kim}}, \bibinfo{author}{\bibfnamefont{C.}~\bibnamefont{Wang}}, \bibinfo{author}{\bibfnamefont{Z.}~\bibnamefont{Hu}}, \bibnamefont{and} \bibinfo{author}{\bibfnamefont{R.}~\bibnamefont{Han}}, \bibinfo{journal}{IEEE Transactions on Terahertz Science and Technology} \textbf{\bibinfo{volume}{9}}, \bibinfo{pages}{349} (\bibinfo{year}{2019}).

\bibitem[{\citenamefont{Yasui et~al.}(2010)\citenamefont{Yasui, Yokoyama, Inaba, Minoshima, Nagatsuma, and Araki}}]{yasui2010terahertz}
\bibinfo{author}{\bibfnamefont{T.}~\bibnamefont{Yasui}}, \bibinfo{author}{\bibfnamefont{S.}~\bibnamefont{Yokoyama}}, \bibinfo{author}{\bibfnamefont{H.}~\bibnamefont{Inaba}}, \bibinfo{author}{\bibfnamefont{K.}~\bibnamefont{Minoshima}}, \bibinfo{author}{\bibfnamefont{T.}~\bibnamefont{Nagatsuma}}, \bibnamefont{and} \bibinfo{author}{\bibfnamefont{T.}~\bibnamefont{Araki}}, \bibinfo{journal}{IEEE Journal of Selected Topics in Quantum Electronics} \textbf{\bibinfo{volume}{17}}, \bibinfo{pages}{191} (\bibinfo{year}{2010}).

\bibitem[{\citenamefont{Consolino et~al.}(2017)\citenamefont{Consolino, Bartalini, and De~Natale}}]{consolino2017terahertz}
\bibinfo{author}{\bibfnamefont{L.}~\bibnamefont{Consolino}}, \bibinfo{author}{\bibfnamefont{S.}~\bibnamefont{Bartalini}}, \bibnamefont{and} \bibinfo{author}{\bibfnamefont{P.}~\bibnamefont{De~Natale}}, \bibinfo{journal}{Journal of Infrared, Millimeter, and Terahertz Waves} \textbf{\bibinfo{volume}{38}}, \bibinfo{pages}{1289} (\bibinfo{year}{2017}).

\bibitem[{\citenamefont{Bellini et~al.}(1992)\citenamefont{Bellini, De~Natale, Di~Lonardo, Fusina, Inguscio, and Prevedelli}}]{bellini1992tunable}
\bibinfo{author}{\bibfnamefont{M.}~\bibnamefont{Bellini}}, \bibinfo{author}{\bibfnamefont{P.}~\bibnamefont{De~Natale}}, \bibinfo{author}{\bibfnamefont{G.}~\bibnamefont{Di~Lonardo}}, \bibinfo{author}{\bibfnamefont{L.}~\bibnamefont{Fusina}}, \bibinfo{author}{\bibfnamefont{M.}~\bibnamefont{Inguscio}}, \bibnamefont{and} \bibinfo{author}{\bibfnamefont{M.}~\bibnamefont{Prevedelli}}, \bibinfo{journal}{Journal of Molecular Spectroscopy} \textbf{\bibinfo{volume}{152}}, \bibinfo{pages}{256} (\bibinfo{year}{1992}).

\bibitem[{\citenamefont{Strumia}(1972)}]{strumia1972proposal}
\bibinfo{author}{\bibfnamefont{F.}~\bibnamefont{Strumia}}, \bibinfo{journal}{Metrologia} \textbf{\bibinfo{volume}{8}}, \bibinfo{pages}{85} (\bibinfo{year}{1972}).

\bibitem[{\citenamefont{Yamamoto et~al.}(2002)\citenamefont{Yamamoto, Takara, and Kawanishi}}]{yamamoto2002generation}
\bibinfo{author}{\bibfnamefont{T.}~\bibnamefont{Yamamoto}}, \bibinfo{author}{\bibfnamefont{H.}~\bibnamefont{Takara}}, \bibnamefont{and} \bibinfo{author}{\bibfnamefont{S.}~\bibnamefont{Kawanishi}}, in \emph{\bibinfo{booktitle}{Proc. Int. Topical Meeting Microwave Photonics}} (\bibinfo{year}{2002}).

\bibitem[{\citenamefont{Champenois et~al.}(2007)\citenamefont{Champenois, Hagel, Houssin, Knoop, Zumsteg, and Vedel}}]{champenois2007thz}
\bibinfo{author}{\bibfnamefont{C.}~\bibnamefont{Champenois}}, \bibinfo{author}{\bibfnamefont{G.}~\bibnamefont{Hagel}}, \bibinfo{author}{\bibfnamefont{M.}~\bibnamefont{Houssin}}, \bibinfo{author}{\bibfnamefont{M.}~\bibnamefont{Knoop}}, \bibinfo{author}{\bibfnamefont{C.}~\bibnamefont{Zumsteg}}, \bibnamefont{and} \bibinfo{author}{\bibfnamefont{F.}~\bibnamefont{Vedel}}, \bibinfo{journal}{Phys. Rev. Lett.} \textbf{\bibinfo{volume}{99}}, \bibinfo{pages}{013001} (\bibinfo{year}{2007}), \urlprefix\url{https://link.aps.org/doi/10.1103/PhysRevLett.99.013001}.

\bibitem[{\citenamefont{Zhou et~al.}(2010)\citenamefont{Zhou, Xu, Chen, and Chen}}]{zhou2010magic}
\bibinfo{author}{\bibfnamefont{X.}~\bibnamefont{Zhou}}, \bibinfo{author}{\bibfnamefont{X.}~\bibnamefont{Xu}}, \bibinfo{author}{\bibfnamefont{X.}~\bibnamefont{Chen}}, \bibnamefont{and} \bibinfo{author}{\bibfnamefont{J.}~\bibnamefont{Chen}}, \bibinfo{journal}{Phys. Rev. A} \textbf{\bibinfo{volume}{81}}, \bibinfo{pages}{012115} (\bibinfo{year}{2010}).

\bibitem[{\citenamefont{Yu et~al.}(2015)\citenamefont{Yu, Geng, Li, Zhou, Duan, Chai, and Yang}}]{yu2015ac}
\bibinfo{author}{\bibfnamefont{G.-H.} \bibnamefont{Yu}}, \bibinfo{author}{\bibfnamefont{Y.-G.} \bibnamefont{Geng}}, \bibinfo{author}{\bibfnamefont{L.}~\bibnamefont{Li}}, \bibinfo{author}{\bibfnamefont{C.}~\bibnamefont{Zhou}}, \bibinfo{author}{\bibfnamefont{C.-B.} \bibnamefont{Duan}}, \bibinfo{author}{\bibfnamefont{R.-P.} \bibnamefont{Chai}}, \bibnamefont{and} \bibinfo{author}{\bibfnamefont{Y.-M.} \bibnamefont{Yang}}, \bibinfo{journal}{Chinese Physics B} \textbf{\bibinfo{volume}{24}}, \bibinfo{pages}{103201} (\bibinfo{year}{2015}).

\bibitem[{\citenamefont{Wang et~al.}(2018)\citenamefont{Wang, Yi, Mawdsley, Kim, Wang, and Han}}]{wang2018chip}
\bibinfo{author}{\bibfnamefont{C.}~\bibnamefont{Wang}}, \bibinfo{author}{\bibfnamefont{X.}~\bibnamefont{Yi}}, \bibinfo{author}{\bibfnamefont{J.}~\bibnamefont{Mawdsley}}, \bibinfo{author}{\bibfnamefont{M.}~\bibnamefont{Kim}}, \bibinfo{author}{\bibfnamefont{Z.}~\bibnamefont{Wang}}, \bibnamefont{and} \bibinfo{author}{\bibfnamefont{R.}~\bibnamefont{Han}}, \bibinfo{journal}{Nature Electronics} \textbf{\bibinfo{volume}{1}}, \bibinfo{pages}{421} (\bibinfo{year}{2018}).

\bibitem[{\citenamefont{Drake et~al.}(2019)\citenamefont{Drake, Briles, Stone, Spencer, Carlson, Hickstein, Li, Westly, Srinivasan, Diddams et~al.}}]{drake2019terahertz}
\bibinfo{author}{\bibfnamefont{T.~E.} \bibnamefont{Drake}}, \bibinfo{author}{\bibfnamefont{T.~C.} \bibnamefont{Briles}}, \bibinfo{author}{\bibfnamefont{J.~R.} \bibnamefont{Stone}}, \bibinfo{author}{\bibfnamefont{D.~T.} \bibnamefont{Spencer}}, \bibinfo{author}{\bibfnamefont{D.~R.} \bibnamefont{Carlson}}, \bibinfo{author}{\bibfnamefont{D.~D.} \bibnamefont{Hickstein}}, \bibinfo{author}{\bibfnamefont{Q.}~\bibnamefont{Li}}, \bibinfo{author}{\bibfnamefont{D.}~\bibnamefont{Westly}}, \bibinfo{author}{\bibfnamefont{K.}~\bibnamefont{Srinivasan}}, \bibinfo{author}{\bibfnamefont{S.~A.} \bibnamefont{Diddams}}, \bibnamefont{et~al.}, \bibinfo{journal}{Phys. Rev. X} \textbf{\bibinfo{volume}{9}}, \bibinfo{pages}{031023} (\bibinfo{year}{2019}).

\bibitem[{\citenamefont{Wang et~al.}(2020)\citenamefont{Wang, Yi, Kim, Yang, and Han}}]{wang2020terahertz}
\bibinfo{author}{\bibfnamefont{C.}~\bibnamefont{Wang}}, \bibinfo{author}{\bibfnamefont{X.}~\bibnamefont{Yi}}, \bibinfo{author}{\bibfnamefont{M.}~\bibnamefont{Kim}}, \bibinfo{author}{\bibfnamefont{Q.~B.} \bibnamefont{Yang}}, \bibnamefont{and} \bibinfo{author}{\bibfnamefont{R.}~\bibnamefont{Han}}, \bibinfo{journal}{IEEE Journal of Solid-State Circuits} \textbf{\bibinfo{volume}{56}}, \bibinfo{pages}{566} (\bibinfo{year}{2020}).

\bibitem[{\citenamefont{Leung et~al.}(2023)\citenamefont{Leung, Iritani, Tiberi, Majewska, Borkowski, Moszynski, and Zelevinsky}}]{leung2023tera}
\bibinfo{author}{\bibfnamefont{K.~H.} \bibnamefont{Leung}}, \bibinfo{author}{\bibfnamefont{B.}~\bibnamefont{Iritani}}, \bibinfo{author}{\bibfnamefont{E.}~\bibnamefont{Tiberi}}, \bibinfo{author}{\bibfnamefont{I.}~\bibnamefont{Majewska}}, \bibinfo{author}{\bibfnamefont{M.}~\bibnamefont{Borkowski}}, \bibinfo{author}{\bibfnamefont{R.}~\bibnamefont{Moszynski}}, \bibnamefont{and} \bibinfo{author}{\bibfnamefont{T.}~\bibnamefont{Zelevinsky}}, \bibinfo{journal}{Phys. Rev. X} \textbf{\bibinfo{volume}{13}}, \bibinfo{pages}{011047} (\bibinfo{year}{2023}), \urlprefix\url{https://link.aps.org/doi/10.1103/PhysRevX.13.011047}.

\bibitem[{\citenamefont{Jyoti et~al.}(2023)\citenamefont{Jyoti, Chakraborty, Yu, Chen, Arora, and Sahoo}}]{jyoti2023zr}
\bibinfo{author}{\bibnamefont{Jyoti}}, \bibinfo{author}{\bibfnamefont{A.}~\bibnamefont{Chakraborty}}, \bibinfo{author}{\bibfnamefont{Y.}~\bibnamefont{Yu}}, \bibinfo{author}{\bibfnamefont{J.}~\bibnamefont{Chen}}, \bibinfo{author}{\bibfnamefont{B.}~\bibnamefont{Arora}}, \bibnamefont{and} \bibinfo{author}{\bibfnamefont{B.~K.} \bibnamefont{Sahoo}}, \bibinfo{journal}{Phys. Rev. A} \textbf{\bibinfo{volume}{108}}, \bibinfo{pages}{023115} (\bibinfo{year}{2023}), \urlprefix\url{https://link.aps.org/doi/10.1103/PhysRevA.108.023115}.

\bibitem[{\citenamefont{~ et~al.}(2021)\citenamefont{~, Kaur, Arora, and Sahoo}}]{jyoti2021spectroscopic}
\bibinfo{author}{\bibfnamefont{J.}~\bibnamefont{~}}, \bibinfo{author}{\bibfnamefont{M.}~\bibnamefont{Kaur}}, \bibinfo{author}{\bibfnamefont{B.}~\bibnamefont{Arora}}, \bibnamefont{and} \bibinfo{author}{\bibfnamefont{B.~K.} \bibnamefont{Sahoo}}, \bibinfo{journal}{Monthly Notices of the Royal Astronomical Society} \textbf{\bibinfo{volume}{507}}, \bibinfo{pages}{4030} (\bibinfo{year}{2021}), ISSN \bibinfo{issn}{0035-8711}, \eprint{https://academic.oup.com/mnras/article-pdf/507/3/4030/40566583/stab2364.pdf}, \urlprefix\url{https://doi.org/10.1093/mnras/stab2364}.

\bibitem[{\citenamefont{Das et~al.}(2017)\citenamefont{Das, Bhowmik, Dutta, and Majumder}}]{das2017electron}
\bibinfo{author}{\bibfnamefont{A.}~\bibnamefont{Das}}, \bibinfo{author}{\bibfnamefont{A.}~\bibnamefont{Bhowmik}}, \bibinfo{author}{\bibfnamefont{N.~N.} \bibnamefont{Dutta}}, \bibnamefont{and} \bibinfo{author}{\bibfnamefont{S.}~\bibnamefont{Majumder}}, \bibinfo{journal}{Journal of Physics B: Atomic, Molecular and Optical Physics} \textbf{\bibinfo{volume}{51}}, \bibinfo{pages}{025001} (\bibinfo{year}{2017}).

\bibitem[{\citenamefont{de~Laeter et~al.}(2003)\citenamefont{de~Laeter, Böhlke, Bièvre, Hidaka, Peiser, Rosman, and Taylor}}]{deLaeter2003abundance}
\bibinfo{author}{\bibfnamefont{J.~R.} \bibnamefont{de~Laeter}}, \bibinfo{author}{\bibfnamefont{J.~K.} \bibnamefont{Böhlke}}, \bibinfo{author}{\bibfnamefont{P.~D.} \bibnamefont{Bièvre}}, \bibinfo{author}{\bibfnamefont{H.}~\bibnamefont{Hidaka}}, \bibinfo{author}{\bibfnamefont{H.~S.} \bibnamefont{Peiser}}, \bibinfo{author}{\bibfnamefont{K.~J.~R.} \bibnamefont{Rosman}}, \bibnamefont{and} \bibinfo{author}{\bibfnamefont{P.~D.~P.} \bibnamefont{Taylor}}, \bibinfo{journal}{Pure and Applied Chemistry} \textbf{\bibinfo{volume}{75}}, \bibinfo{pages}{683} (\bibinfo{year}{2003}), \urlprefix\url{https://doi.org/10.1351/pac200375060683}.

\bibitem[{\citenamefont{Silver et~al.}(1994)\citenamefont{Silver, Varney, Margolis, Baird, Grant, Groves, Hallett, Handford, Hirst, Holmes et~al.}}]{silver1994oxford}
\bibinfo{author}{\bibfnamefont{J.}~\bibnamefont{Silver}}, \bibinfo{author}{\bibfnamefont{A.}~\bibnamefont{Varney}}, \bibinfo{author}{\bibfnamefont{H.}~\bibnamefont{Margolis}}, \bibinfo{author}{\bibfnamefont{P.}~\bibnamefont{Baird}}, \bibinfo{author}{\bibfnamefont{I.}~\bibnamefont{Grant}}, \bibinfo{author}{\bibfnamefont{P.}~\bibnamefont{Groves}}, \bibinfo{author}{\bibfnamefont{W.}~\bibnamefont{Hallett}}, \bibinfo{author}{\bibfnamefont{A.}~\bibnamefont{Handford}}, \bibinfo{author}{\bibfnamefont{P.}~\bibnamefont{Hirst}}, \bibinfo{author}{\bibfnamefont{A.}~\bibnamefont{Holmes}}, \bibnamefont{et~al.}, \bibinfo{journal}{Review of scientific instruments} \textbf{\bibinfo{volume}{65}}, \bibinfo{pages}{1072} (\bibinfo{year}{1994}).

\bibitem[{\citenamefont{Nakamura et~al.}(2008)\citenamefont{Nakamura, Kikuchi, Sakaue, and Watanabe}}]{nakamura2008compact}
\bibinfo{author}{\bibfnamefont{N.}~\bibnamefont{Nakamura}}, \bibinfo{author}{\bibfnamefont{H.}~\bibnamefont{Kikuchi}}, \bibinfo{author}{\bibfnamefont{H.~A.} \bibnamefont{Sakaue}}, \bibnamefont{and} \bibinfo{author}{\bibfnamefont{T.}~\bibnamefont{Watanabe}}, \bibinfo{journal}{Review of Scientific Instruments} \textbf{\bibinfo{volume}{79}}, \bibinfo{pages}{063104} (\bibinfo{year}{2008}).

\bibitem[{\citenamefont{Agnihotri et~al.}(2011)\citenamefont{Agnihotri, Kelkar, Kasthurirangan, Thulasiram, Desai, Fernandez, and Tribedi}}]{agnihotri2011ecr}
\bibinfo{author}{\bibfnamefont{A.}~\bibnamefont{Agnihotri}}, \bibinfo{author}{\bibfnamefont{A.}~\bibnamefont{Kelkar}}, \bibinfo{author}{\bibfnamefont{S.}~\bibnamefont{Kasthurirangan}}, \bibinfo{author}{\bibfnamefont{K.}~\bibnamefont{Thulasiram}}, \bibinfo{author}{\bibfnamefont{C.}~\bibnamefont{Desai}}, \bibinfo{author}{\bibfnamefont{W.}~\bibnamefont{Fernandez}}, \bibnamefont{and} \bibinfo{author}{\bibfnamefont{L.}~\bibnamefont{Tribedi}}, \bibinfo{journal}{Physica Scripta} \textbf{\bibinfo{volume}{2011}}, \bibinfo{pages}{014038} (\bibinfo{year}{2011}).

\bibitem[{\citenamefont{Chou et~al.}(2010)\citenamefont{Chou, Hume, Koelemeij, Wineland, and Rosenband}}]{chou2010al+}
\bibinfo{author}{\bibfnamefont{C.~W.} \bibnamefont{Chou}}, \bibinfo{author}{\bibfnamefont{D.~B.} \bibnamefont{Hume}}, \bibinfo{author}{\bibfnamefont{J.~C.~J.} \bibnamefont{Koelemeij}}, \bibinfo{author}{\bibfnamefont{D.~J.} \bibnamefont{Wineland}}, \bibnamefont{and} \bibinfo{author}{\bibfnamefont{T.}~\bibnamefont{Rosenband}}, \bibinfo{journal}{Phys. Rev. Lett.} \textbf{\bibinfo{volume}{104}}, \bibinfo{pages}{070802} (\bibinfo{year}{2010}), \urlprefix\url{https://link.aps.org/doi/10.1103/PhysRevLett.104.070802}.

\bibitem[{\citenamefont{Hannig et~al.}(2019)\citenamefont{Hannig, Pelzer, Scharnhorst, Kramer, Stepanova, Xu, Spethmann, Leroux, Mehlstäubler, and Schmidt}}]{hannig2019al}
\bibinfo{author}{\bibfnamefont{S.}~\bibnamefont{Hannig}}, \bibinfo{author}{\bibfnamefont{L.}~\bibnamefont{Pelzer}}, \bibinfo{author}{\bibfnamefont{N.}~\bibnamefont{Scharnhorst}}, \bibinfo{author}{\bibfnamefont{J.}~\bibnamefont{Kramer}}, \bibinfo{author}{\bibfnamefont{M.}~\bibnamefont{Stepanova}}, \bibinfo{author}{\bibfnamefont{Z.~T.} \bibnamefont{Xu}}, \bibinfo{author}{\bibfnamefont{N.}~\bibnamefont{Spethmann}}, \bibinfo{author}{\bibfnamefont{I.~D.} \bibnamefont{Leroux}}, \bibinfo{author}{\bibfnamefont{T.~E.} \bibnamefont{Mehlstäubler}}, \bibnamefont{and} \bibinfo{author}{\bibfnamefont{P.~O.} \bibnamefont{Schmidt}}, \bibinfo{journal}{Review of Scientific Instruments} \textbf{\bibinfo{volume}{90}} (\bibinfo{year}{2019}), ISSN \bibinfo{issn}{0034-6748}, \bibinfo{note}{053204}, \eprint{https://pubs.aip.org/aip/rsi/article-pdf/doi/10.1063/1.5090583/14695430/053204\_1\_online.pdf}, \urlprefix\url{https://doi.org/10.1063/1.5090583}.

\bibitem[{\citenamefont{Micalizio et~al.}(2021)\citenamefont{Micalizio, Levi, Calosso, Gozzelino, and Godone}}]{micalizio2021pulsed}
\bibinfo{author}{\bibfnamefont{S.}~\bibnamefont{Micalizio}}, \bibinfo{author}{\bibfnamefont{F.}~\bibnamefont{Levi}}, \bibinfo{author}{\bibfnamefont{C.}~\bibnamefont{Calosso}}, \bibinfo{author}{\bibfnamefont{M.}~\bibnamefont{Gozzelino}}, \bibnamefont{and} \bibinfo{author}{\bibfnamefont{A.}~\bibnamefont{Godone}}, \bibinfo{journal}{GPS Solutions} \textbf{\bibinfo{volume}{25}}, \bibinfo{pages}{94} (\bibinfo{year}{2021}).

\bibitem[{\citenamefont{Wen et~al.}(2020)\citenamefont{Wen, Zong, Zhang, Yang, Wang, Zhang, Bo, Peng, Cui, Xu et~al.}}]{Wen_2020}
\bibinfo{author}{\bibfnamefont{N.}~\bibnamefont{Wen}}, \bibinfo{author}{\bibfnamefont{N.}~\bibnamefont{Zong}}, \bibinfo{author}{\bibfnamefont{F.-F.} \bibnamefont{Zhang}}, \bibinfo{author}{\bibfnamefont{F.}~\bibnamefont{Yang}}, \bibinfo{author}{\bibfnamefont{Z.-M.} \bibnamefont{Wang}}, \bibinfo{author}{\bibfnamefont{S.-J.} \bibnamefont{Zhang}}, \bibinfo{author}{\bibfnamefont{Y.}~\bibnamefont{Bo}}, \bibinfo{author}{\bibfnamefont{Q.-J.} \bibnamefont{Peng}}, \bibinfo{author}{\bibfnamefont{D.-F.} \bibnamefont{Cui}}, \bibinfo{author}{\bibfnamefont{Z.-Y.} \bibnamefont{Xu}}, \bibnamefont{et~al.}, \bibinfo{journal}{Laser Physics Letters} \textbf{\bibinfo{volume}{17}}, \bibinfo{pages}{105001} (\bibinfo{year}{2020}), \urlprefix\url{https://dx.doi.org/10.1088/1612-202X/abac13}.

\bibitem[{\citenamefont{Singh et~al.}(2024)\citenamefont{Singh, Fareed, Shirinabadi, Marcelino, Zhu, Légaré, and Ozaki}}]{singh2024hhg}
\bibinfo{author}{\bibfnamefont{M.}~\bibnamefont{Singh}}, \bibinfo{author}{\bibfnamefont{M.~A.} \bibnamefont{Fareed}}, \bibinfo{author}{\bibfnamefont{R.~G.} \bibnamefont{Shirinabadi}}, \bibinfo{author}{\bibfnamefont{R.}~\bibnamefont{Marcelino}}, \bibinfo{author}{\bibfnamefont{F.}~\bibnamefont{Zhu}}, \bibinfo{author}{\bibfnamefont{F.}~\bibnamefont{Légaré}}, \bibnamefont{and} \bibinfo{author}{\bibfnamefont{T.}~\bibnamefont{Ozaki}}, \bibinfo{journal}{Fundamental Plasma Physics} \textbf{\bibinfo{volume}{10}}, \bibinfo{pages}{100043} (\bibinfo{year}{2024}), ISSN \bibinfo{issn}{2772-8285}, \urlprefix\url{https://www.sciencedirect.com/science/article/pii/S2772828524000086}.

\bibitem[{\citenamefont{Merkel et~al.}(2019)\citenamefont{Merkel, Thirumalai, Tarlton, Sch{\"a}fer, Ballance, Harty, and Lucas}}]{merkel2019magnetic}
\bibinfo{author}{\bibfnamefont{B.}~\bibnamefont{Merkel}}, \bibinfo{author}{\bibfnamefont{K.}~\bibnamefont{Thirumalai}}, \bibinfo{author}{\bibfnamefont{J.}~\bibnamefont{Tarlton}}, \bibinfo{author}{\bibfnamefont{V.}~\bibnamefont{Sch{\"a}fer}}, \bibinfo{author}{\bibfnamefont{C.}~\bibnamefont{Ballance}}, \bibinfo{author}{\bibfnamefont{T.}~\bibnamefont{Harty}}, \bibnamefont{and} \bibinfo{author}{\bibfnamefont{D.}~\bibnamefont{Lucas}}, \bibinfo{journal}{Review of Scientific Instruments} \textbf{\bibinfo{volume}{90}}, \bibinfo{pages}{044702} (\bibinfo{year}{2019}).

\bibitem[{\citenamefont{Blundell et~al.}(1991)\citenamefont{Blundell, Johnson, and Sapirstein}}]{blundell1991relativistic}
\bibinfo{author}{\bibfnamefont{S.~A.} \bibnamefont{Blundell}}, \bibinfo{author}{\bibfnamefont{W.~R.} \bibnamefont{Johnson}}, \bibnamefont{and} \bibinfo{author}{\bibfnamefont{J.}~\bibnamefont{Sapirstein}}, \bibinfo{journal}{Phys. Rev. A} \textbf{\bibinfo{volume}{43}}, \bibinfo{pages}{3407} (\bibinfo{year}{1991}), \urlprefix\url{https://link.aps.org/doi/10.1103/PhysRevA.43.3407}.

\bibitem[{\citenamefont{(Ilyabaev) et~al.}(1994)\citenamefont{(Ilyabaev), Kaldor, and Ishikawa}}]{ILYABAEV199482}
\bibinfo{author}{\bibfnamefont{E.~E.} \bibnamefont{(Ilyabaev)}}, \bibinfo{author}{\bibfnamefont{U.}~\bibnamefont{Kaldor}}, \bibnamefont{and} \bibinfo{author}{\bibfnamefont{Y.}~\bibnamefont{Ishikawa}}, \bibinfo{journal}{Chem. Phys. Lett.} \textbf{\bibinfo{volume}{222}}, \bibinfo{pages}{82} (\bibinfo{year}{1994}), ISSN \bibinfo{issn}{0009-2614}, \urlprefix\url{https://www.sciencedirect.com/science/article/pii/0009261494003173}.

\bibitem[{\citenamefont{Lindroth and Ynnerman}(1993)}]{lindroth1993ab}
\bibinfo{author}{\bibfnamefont{E.}~\bibnamefont{Lindroth}} \bibnamefont{and} \bibinfo{author}{\bibfnamefont{A.}~\bibnamefont{Ynnerman}}, \bibinfo{journal}{Phys. Rev. A} \textbf{\bibinfo{volume}{47}}, \bibinfo{pages}{961} (\bibinfo{year}{1993}), \urlprefix\url{https://link.aps.org/doi/10.1103/PhysRevA.47.961}.

\bibitem[{\citenamefont{Sahoo et~al.}(2004)\citenamefont{Sahoo, Majumder, Chaudhuri, Das, and Mukherjee}}]{sahoo2004ab}
\bibinfo{author}{\bibfnamefont{B.~K.} \bibnamefont{Sahoo}}, \bibinfo{author}{\bibfnamefont{S.}~\bibnamefont{Majumder}}, \bibinfo{author}{\bibfnamefont{R.~K.} \bibnamefont{Chaudhuri}}, \bibinfo{author}{\bibfnamefont{B.}~\bibnamefont{Das}}, \bibnamefont{and} \bibinfo{author}{\bibfnamefont{D.}~\bibnamefont{Mukherjee}}, \bibinfo{journal}{Journal of Physics B: Atomic, Molecular and Optical Physics} \textbf{\bibinfo{volume}{37}}, \bibinfo{pages}{3409} (\bibinfo{year}{2004}).

\bibitem[{\citenamefont{Nandy and Sahoo}(2014)}]{nandy2014quadrupole}
\bibinfo{author}{\bibfnamefont{D.~K.} \bibnamefont{Nandy}} \bibnamefont{and} \bibinfo{author}{\bibfnamefont{B.~K.} \bibnamefont{Sahoo}}, \bibinfo{journal}{Phys. Rev. A} \textbf{\bibinfo{volume}{90}}, \bibinfo{pages}{050503} (\bibinfo{year}{2014}), \urlprefix\url{https://link.aps.org/doi/10.1103/PhysRevA.90.050503}.

\bibitem[{\citenamefont{Ginges and Berengut}(2016)}]{ginges2016atomic}
\bibinfo{author}{\bibfnamefont{J.~S.~M.} \bibnamefont{Ginges}} \bibnamefont{and} \bibinfo{author}{\bibfnamefont{J.~C.} \bibnamefont{Berengut}}, \bibinfo{journal}{Phys. Rev. A} \textbf{\bibinfo{volume}{93}}, \bibinfo{pages}{052509} (\bibinfo{year}{2016}), \urlprefix\url{https://link.aps.org/doi/10.1103/PhysRevA.93.052509}.

\bibitem[{\citenamefont{Sahoo}(2016)}]{sahoo2016conform}
\bibinfo{author}{\bibfnamefont{B.~K.} \bibnamefont{Sahoo}}, \bibinfo{journal}{Phys. Rev. A} \textbf{\bibinfo{volume}{93}}, \bibinfo{pages}{022503} (\bibinfo{year}{2016}), \urlprefix\url{https://link.aps.org/doi/10.1103/PhysRevA.93.022503}.

\bibitem[{\citenamefont{Flambaum and Ginges}(2005)}]{flambaum2005radiative}
\bibinfo{author}{\bibfnamefont{V.~V.} \bibnamefont{Flambaum}} \bibnamefont{and} \bibinfo{author}{\bibfnamefont{J.~S.~M.} \bibnamefont{Ginges}}, \bibinfo{journal}{Phys. Rev. A} \textbf{\bibinfo{volume}{72}}, \bibinfo{pages}{052115} (\bibinfo{year}{2005}), \urlprefix\url{https://link.aps.org/doi/10.1103/PhysRevA.72.052115}.

\bibitem[{\citenamefont{Li et~al.}(2018)\citenamefont{Li, Yu, and Sahoo}}]{li2018cc}
\bibinfo{author}{\bibfnamefont{C.-B.} \bibnamefont{Li}}, \bibinfo{author}{\bibfnamefont{Y.}~\bibnamefont{Yu}}, \bibnamefont{and} \bibinfo{author}{\bibfnamefont{B.~K.} \bibnamefont{Sahoo}}, \bibinfo{journal}{Phys. Rev. A} \textbf{\bibinfo{volume}{97}}, \bibinfo{pages}{022512} (\bibinfo{year}{2018}), \urlprefix\url{https://link.aps.org/doi/10.1103/PhysRevA.97.022512}.

\bibitem[{\citenamefont{{\v{C}}{\'\i}{\v{z}}ek}(1969)}]{vcivzek1969use}
\bibinfo{author}{\bibfnamefont{J.}~\bibnamefont{{\v{C}}{\'\i}{\v{z}}ek}}, \bibinfo{journal}{Adv. Chem. Phys.} pp. \bibinfo{pages}{35--89} (\bibinfo{year}{1969}).

\bibitem[{\citenamefont{Lindgren and Morrison}(2012)}]{lindgren2012atomic}
\bibinfo{author}{\bibfnamefont{I.}~\bibnamefont{Lindgren}} \bibnamefont{and} \bibinfo{author}{\bibfnamefont{J.}~\bibnamefont{Morrison}}, \emph{\bibinfo{title}{Atomic many-body theory}}, vol.~\bibinfo{volume}{3} (\bibinfo{publisher}{Springer Science \& Business Media}, \bibinfo{year}{2012}).

\bibitem[{\citenamefont{Sahoo}(2005)}]{bijaya2005cc}
\bibinfo{author}{\bibfnamefont{B.~K.} \bibnamefont{Sahoo}}, \bibinfo{journal}{Ph.D. Thesis}  (\bibinfo{year}{2005}).

\bibitem[{\citenamefont{Boys}(1950)}]{boys1950electronic}
\bibinfo{author}{\bibfnamefont{S.~F.} \bibnamefont{Boys}}, \bibinfo{journal}{Proceedings of the Royal Society of London. Series A. Mathematical and Physical Sciences} \textbf{\bibinfo{volume}{200}}, \bibinfo{pages}{542} (\bibinfo{year}{1950}).

\bibitem[{\citenamefont{Wilson}(1997)}]{wilson1997practical}
\bibinfo{author}{\bibfnamefont{S.}~\bibnamefont{Wilson}}, in \emph{\bibinfo{booktitle}{Problem Solving in Computational Molecular Science: Molecules in Different Environments}} (\bibinfo{publisher}{Springer}, \bibinfo{year}{1997}), pp. \bibinfo{pages}{109--158}.

\bibitem[{\citenamefont{Mohanty and Clementi}(1989)}]{mohanty1989kinetically}
\bibinfo{author}{\bibfnamefont{A.~K.} \bibnamefont{Mohanty}} \bibnamefont{and} \bibinfo{author}{\bibfnamefont{E.}~\bibnamefont{Clementi}}, \bibinfo{journal}{Chemical physics letters} \textbf{\bibinfo{volume}{157}}, \bibinfo{pages}{348} (\bibinfo{year}{1989}).

\bibitem[{\citenamefont{Kramida et~al.}(2020)\citenamefont{Kramida, {Yu.~Ralchenko}, Reader, and { NIST ASD Team}}}]{ralchenko2008nist}
\bibinfo{author}{\bibfnamefont{A.}~\bibnamefont{Kramida}}, \bibinfo{author}{\bibnamefont{{Yu.~Ralchenko}}}, \bibinfo{author}{\bibfnamefont{J.}~\bibnamefont{Reader}}, \bibnamefont{and} \bibinfo{author}{\bibnamefont{{ NIST ASD Team}}}, \bibinfo{howpublished}{{NIST Atomic Spectra Database (ver. 5.8), [Online]. Available: {\tt{https://physics.nist.gov/asd}} [2021, February 24]. National Institute of Standards and Technology, Gaithersburg, MD.}} (\bibinfo{year}{2020}).

\bibitem[{\citenamefont{Manakov et~al.}(1986)\citenamefont{Manakov, Ovsiannikov, and Rapoport}}]{manakov1986atoms}
\bibinfo{author}{\bibfnamefont{N.~L.} \bibnamefont{Manakov}}, \bibinfo{author}{\bibfnamefont{V.~D.} \bibnamefont{Ovsiannikov}}, \bibnamefont{and} \bibinfo{author}{\bibfnamefont{L.~P.} \bibnamefont{Rapoport}}, \bibinfo{journal}{Physics Reports} \textbf{\bibinfo{volume}{141}}, \bibinfo{pages}{320} (\bibinfo{year}{1986}), \urlprefix\url{https://www.sciencedirect.com/science/article/abs/pii/S0370157386800011}.

\bibitem[{\citenamefont{Singh et~al.}(2016)\citenamefont{Singh, Kaur, Sahoo, and Arora}}]{singh2016comparing}
\bibinfo{author}{\bibfnamefont{S.}~\bibnamefont{Singh}}, \bibinfo{author}{\bibfnamefont{K.}~\bibnamefont{Kaur}}, \bibinfo{author}{\bibfnamefont{B.}~\bibnamefont{Sahoo}}, \bibnamefont{and} \bibinfo{author}{\bibfnamefont{B.}~\bibnamefont{Arora}}, \bibinfo{journal}{Journal of Physics B: Atomic, Molecular and Optical Physics} \textbf{\bibinfo{volume}{49}}, \bibinfo{pages}{145005} (\bibinfo{year}{2016}), \urlprefix\url{https://iopscience.iop.org/article/10.1088/0953-4075/49/14/145005/meta}.

\bibitem[{\citenamefont{Schäffer et~al.}(2017)\citenamefont{Schäffer, Christensen, Rathmann, Appel, Henriksen, and Thomsen}}]{schaffer2017towards}
\bibinfo{author}{\bibfnamefont{S.~A.} \bibnamefont{Schäffer}}, \bibinfo{author}{\bibfnamefont{B.~T.~R.} \bibnamefont{Christensen}}, \bibinfo{author}{\bibfnamefont{S.~M.} \bibnamefont{Rathmann}}, \bibinfo{author}{\bibfnamefont{M.~H.} \bibnamefont{Appel}}, \bibinfo{author}{\bibfnamefont{M.~R.} \bibnamefont{Henriksen}}, \bibnamefont{and} \bibinfo{author}{\bibfnamefont{J.~W.} \bibnamefont{Thomsen}} (\bibinfo{publisher}{IOP Publishing}, \bibinfo{year}{2017}), vol. \bibinfo{volume}{810}, p. \bibinfo{pages}{012002}, \urlprefix\url{https://dx.doi.org/10.1088/1742-6596/810/1/012002}.

\bibitem[{\citenamefont{Bonin and Kresin}(1997)}]{bonin1997dipole}
\bibinfo{author}{\bibfnamefont{K.~D.} \bibnamefont{Bonin}} \bibnamefont{and} \bibinfo{author}{\bibfnamefont{V.~V.} \bibnamefont{Kresin}}, \emph{\bibinfo{title}{Electric-Dipole Polarizabilities of Atoms, Molecules, and Clusters}} (\bibinfo{publisher}{WORLD SCIENTIFIC}, \bibinfo{year}{1997}).

\bibitem[{\citenamefont{Kaur et~al.}(2015)\citenamefont{Kaur, Nandy, Arora, and Sahoo}}]{kaur2015properties}
\bibinfo{author}{\bibfnamefont{J.}~\bibnamefont{Kaur}}, \bibinfo{author}{\bibfnamefont{D.~K.} \bibnamefont{Nandy}}, \bibinfo{author}{\bibfnamefont{B.}~\bibnamefont{Arora}}, \bibnamefont{and} \bibinfo{author}{\bibfnamefont{B.~K.} \bibnamefont{Sahoo}}, \bibinfo{journal}{Phys. Rev. A} \textbf{\bibinfo{volume}{91}}, \bibinfo{pages}{012705} (\bibinfo{year}{2015}), \urlprefix\url{https://link.aps.org/doi/10.1103/PhysRevA.91.012705}.

\bibitem[{\citenamefont{Safronova et~al.}(2006)\citenamefont{Safronova, Safronova, and Beiersdorfer}}]{safronova2006multipole}
\bibinfo{author}{\bibfnamefont{U.}~\bibnamefont{Safronova}}, \bibinfo{author}{\bibfnamefont{A.}~\bibnamefont{Safronova}}, \bibnamefont{and} \bibinfo{author}{\bibfnamefont{P.}~\bibnamefont{Beiersdorfer}}, \bibinfo{journal}{Journal of Physics B: Atomic, Molecular and Optical Physics} \textbf{\bibinfo{volume}{39}}, \bibinfo{pages}{4491} (\bibinfo{year}{2006}).

\bibitem[{\citenamefont{Johnson et~al.}(1995)\citenamefont{Johnson, Plante, and Sapirstein}}]{johnson1995relativistic}
\bibinfo{author}{\bibfnamefont{W.}~\bibnamefont{Johnson}}, \bibinfo{author}{\bibfnamefont{D.}~\bibnamefont{Plante}}, \bibnamefont{and} \bibinfo{author}{\bibfnamefont{J.}~\bibnamefont{Sapirstein}}, in \emph{\bibinfo{booktitle}{Advances in Atomic, Molecular, and Optical Physics}} (\bibinfo{publisher}{Elsevier}, \bibinfo{year}{1995}), vol.~\bibinfo{volume}{35}, pp. \bibinfo{pages}{255--329}.

\bibitem[{\citenamefont{Pan et~al.}(2020)\citenamefont{Pan, Arora, Yu, Sahoo, and Chen}}]{pan2020optical}
\bibinfo{author}{\bibfnamefont{D.}~\bibnamefont{Pan}}, \bibinfo{author}{\bibfnamefont{B.}~\bibnamefont{Arora}}, \bibinfo{author}{\bibfnamefont{Y.}~\bibnamefont{Yu}}, \bibinfo{author}{\bibfnamefont{B.~K.} \bibnamefont{Sahoo}}, \bibnamefont{and} \bibinfo{author}{\bibfnamefont{J.}~\bibnamefont{Chen}}, \bibinfo{journal}{Phys. Rev. A} \textbf{\bibinfo{volume}{102}}, \bibinfo{pages}{041101} (\bibinfo{year}{2020}), \urlprefix\url{https://link.aps.org/doi/10.1103/PhysRevA.102.041101}.

\bibitem[{\citenamefont{Sahoo et~al.}(2015)\citenamefont{Sahoo, Nandy, Das, and Sakemi}}]{sahoo2015correlation}
\bibinfo{author}{\bibfnamefont{B.~K.} \bibnamefont{Sahoo}}, \bibinfo{author}{\bibfnamefont{D.~K.} \bibnamefont{Nandy}}, \bibinfo{author}{\bibfnamefont{B.~P.} \bibnamefont{Das}}, \bibnamefont{and} \bibinfo{author}{\bibfnamefont{Y.}~\bibnamefont{Sakemi}}, \bibinfo{journal}{Phys. Rev. A} \textbf{\bibinfo{volume}{91}}, \bibinfo{pages}{042507} (\bibinfo{year}{2015}), \urlprefix\url{https://link.aps.org/doi/10.1103/PhysRevA.91.042507}.

\bibitem[{\citenamefont{Schwartz}(1955)}]{schwartz1955hfs}
\bibinfo{author}{\bibfnamefont{C.}~\bibnamefont{Schwartz}}, \bibinfo{journal}{Phys. Rev.} \textbf{\bibinfo{volume}{97}}, \bibinfo{pages}{380} (\bibinfo{year}{1955}), \urlprefix\url{https://link.aps.org/doi/10.1103/PhysRev.97.380}.

\bibitem[{\citenamefont{Sobelman}(1979)}]{sobelman1979angular}
\bibinfo{author}{\bibfnamefont{I.~I.} \bibnamefont{Sobelman}}, in \emph{\bibinfo{booktitle}{Atomic Spectra and Radiative Transitions}} (\bibinfo{publisher}{Springer}, \bibinfo{year}{1979}), pp. \bibinfo{pages}{53--88}, \urlprefix\url{https://link.springer.com/chapter/10.1007/978-3-662-04589-3_4}.

\bibitem[{\citenamefont{Indelicato}(1996)}]{indelicato1995neg}
\bibinfo{author}{\bibfnamefont{P.}~\bibnamefont{Indelicato}}, \bibinfo{journal}{Phys. Rev. Lett.} \textbf{\bibinfo{volume}{77}}, \bibinfo{pages}{3323} (\bibinfo{year}{1996}), \urlprefix\url{https://link.aps.org/doi/10.1103/PhysRevLett.77.3323}.

\bibitem[{\citenamefont{Farley and Wing}(1981)}]{farley1981accurate}
\bibinfo{author}{\bibfnamefont{J.~W.} \bibnamefont{Farley}} \bibnamefont{and} \bibinfo{author}{\bibfnamefont{W.~H.} \bibnamefont{Wing}}, \bibinfo{journal}{Phys. Rev. A} \textbf{\bibinfo{volume}{23}}, \bibinfo{pages}{2397} (\bibinfo{year}{1981}), \urlprefix\url{https://link.aps.org/doi/10.1103/PhysRevA.23.2397}.

\bibitem[{\citenamefont{Yu and Sahoo}(2018)}]{yu2018selected}
\bibinfo{author}{\bibfnamefont{Y.-m.} \bibnamefont{Yu}} \bibnamefont{and} \bibinfo{author}{\bibfnamefont{B.~K.} \bibnamefont{Sahoo}}, \bibinfo{journal}{Phys. Rev. A} \textbf{\bibinfo{volume}{97}}, \bibinfo{pages}{041403} (\bibinfo{year}{2018}), \urlprefix\url{https://link.aps.org/doi/10.1103/PhysRevA.97.041403}.

\bibitem[{\citenamefont{Arora et~al.}(2007)\citenamefont{Arora, Safronova, and Clark}}]{arora2007blackbody}
\bibinfo{author}{\bibfnamefont{B.}~\bibnamefont{Arora}}, \bibinfo{author}{\bibfnamefont{M.}~\bibnamefont{Safronova}}, \bibnamefont{and} \bibinfo{author}{\bibfnamefont{C.~W.} \bibnamefont{Clark}}, \bibinfo{journal}{Phys. Rev. A} \textbf{\bibinfo{volume}{76}}, \bibinfo{pages}{064501} (\bibinfo{year}{2007}), \urlprefix\url{https://journals.aps.org/pra/abstract/10.1103/PhysRevA.76.064501}.

\bibitem[{\citenamefont{Arora et~al.}(2012)\citenamefont{Arora, Nandy, and Sahoo}}]{arora2012multipolar}
\bibinfo{author}{\bibfnamefont{B.}~\bibnamefont{Arora}}, \bibinfo{author}{\bibfnamefont{D.~K.} \bibnamefont{Nandy}}, \bibnamefont{and} \bibinfo{author}{\bibfnamefont{B.~K.} \bibnamefont{Sahoo}}, \bibinfo{journal}{Phys. Rev. A} \textbf{\bibinfo{volume}{85}}, \bibinfo{pages}{012506} (\bibinfo{year}{2012}), \urlprefix\url{https://link.aps.org/doi/10.1103/PhysRevA.85.012506}.

\bibitem[{\citenamefont{Campbell et~al.}(2012)\citenamefont{Campbell, Radnaev, Kuzmich, Dzuba, Flambaum, and Derevianko}}]{campbell2012single}
\bibinfo{author}{\bibfnamefont{C.~J.} \bibnamefont{Campbell}}, \bibinfo{author}{\bibfnamefont{A.~G.} \bibnamefont{Radnaev}}, \bibinfo{author}{\bibfnamefont{A.}~\bibnamefont{Kuzmich}}, \bibinfo{author}{\bibfnamefont{V.~A.} \bibnamefont{Dzuba}}, \bibinfo{author}{\bibfnamefont{V.~V.} \bibnamefont{Flambaum}}, \bibnamefont{and} \bibinfo{author}{\bibfnamefont{A.}~\bibnamefont{Derevianko}}, \bibinfo{journal}{Phys. Rev. Lett.} \textbf{\bibinfo{volume}{108}}, \bibinfo{pages}{120802} (\bibinfo{year}{2012}), \urlprefix\url{https://journals.aps.org/prl/abstract/10.1103/PhysRevLett.108.120802}.

\bibitem[{\citenamefont{Dzuba et~al.}(2021)\citenamefont{Dzuba, Allehabi, Flambaum, Li, and Schiller}}]{dzuba2021time}
\bibinfo{author}{\bibfnamefont{V.}~\bibnamefont{Dzuba}}, \bibinfo{author}{\bibfnamefont{S.~O.} \bibnamefont{Allehabi}}, \bibinfo{author}{\bibfnamefont{V.}~\bibnamefont{Flambaum}}, \bibinfo{author}{\bibfnamefont{J.}~\bibnamefont{Li}}, \bibnamefont{and} \bibinfo{author}{\bibfnamefont{S.}~\bibnamefont{Schiller}}, \bibinfo{journal}{Phys. Rev. A} \textbf{\bibinfo{volume}{103}}, \bibinfo{pages}{022822} (\bibinfo{year}{2021}), \urlprefix\url{https://journals.aps.org/pra/abstract/10.1103/PhysRevA.103.022822}.

\bibitem[{\citenamefont{Takamoto et~al.}(2006)\citenamefont{Takamoto, Hong, Higashi, Fujii, Imae, and Katori}}]{takamoto2006improved}
\bibinfo{author}{\bibfnamefont{M.}~\bibnamefont{Takamoto}}, \bibinfo{author}{\bibfnamefont{F.-L.} \bibnamefont{Hong}}, \bibinfo{author}{\bibfnamefont{R.}~\bibnamefont{Higashi}}, \bibinfo{author}{\bibfnamefont{Y.}~\bibnamefont{Fujii}}, \bibinfo{author}{\bibfnamefont{M.}~\bibnamefont{Imae}}, \bibnamefont{and} \bibinfo{author}{\bibfnamefont{H.}~\bibnamefont{Katori}}, \bibinfo{journal}{Journal of the Physical Society of Japan} \textbf{\bibinfo{volume}{75}}, \bibinfo{pages}{104302} (\bibinfo{year}{2006}), \urlprefix\url{https://journals.jps.jp/doi/abs/10.1143/JPSJ.75.104302}.

\bibitem[{\citenamefont{Porsev and Safronova}(2020)}]{porsev2020calculation}
\bibinfo{author}{\bibfnamefont{S.~G.} \bibnamefont{Porsev}} \bibnamefont{and} \bibinfo{author}{\bibfnamefont{M.~S.} \bibnamefont{Safronova}}, \bibinfo{journal}{Phys. Rev. A} \textbf{\bibinfo{volume}{102}}, \bibinfo{pages}{012811} (\bibinfo{year}{2020}), \urlprefix\url{https://link.aps.org/doi/10.1103/PhysRevA.102.012811}.

\bibitem[{\citenamefont{Itano}(2000)}]{itano2000external}
\bibinfo{author}{\bibfnamefont{W.~M.} \bibnamefont{Itano}}, \bibinfo{journal}{Journal of research of the National Institute of Standards and Technology} \textbf{\bibinfo{volume}{105}}, \bibinfo{pages}{829} (\bibinfo{year}{2000}), \urlprefix\url{https://www.ncbi.nlm.nih.gov/pmc/articles/PMC4877145/}.

\bibitem[{\citenamefont{Porsev et~al.}(2020)\citenamefont{Porsev, Safronova, Safronova, Schmidt, Bondarev, Kozlov, Tupitsyn, and Cheung}}]{porsev2020optical}
\bibinfo{author}{\bibfnamefont{S.}~\bibnamefont{Porsev}}, \bibinfo{author}{\bibfnamefont{U.}~\bibnamefont{Safronova}}, \bibinfo{author}{\bibfnamefont{M.}~\bibnamefont{Safronova}}, \bibinfo{author}{\bibfnamefont{P.}~\bibnamefont{Schmidt}}, \bibinfo{author}{\bibfnamefont{A.}~\bibnamefont{Bondarev}}, \bibinfo{author}{\bibfnamefont{M.}~\bibnamefont{Kozlov}}, \bibinfo{author}{\bibfnamefont{I.}~\bibnamefont{Tupitsyn}}, \bibnamefont{and} \bibinfo{author}{\bibfnamefont{C.}~\bibnamefont{Cheung}}, \bibinfo{journal}{Phys. Rev. A} \textbf{\bibinfo{volume}{102}}, \bibinfo{pages}{012802} (\bibinfo{year}{2020}), \urlprefix\url{https://journals.aps.org/pra/abstract/10.1103/PhysRevA.102.012802}.

\bibitem[{\citenamefont{Sahoo}(2015)}]{sahoo2015springer}
\bibinfo{author}{\bibfnamefont{B.~K.} \bibnamefont{Sahoo}}, \bibinfo{journal}{Handbook of Relativistic Quantum Chemistry; Liu, W., Ed.; Springer: Berlin/Heidelberg, Germany} pp. \bibinfo{pages}{611--655} (\bibinfo{year}{2015}).

\bibitem[{\citenamefont{Gu\'ena et~al.}(2011)\citenamefont{Gu\'ena, Li, Gibble, Bize, and Clairon}}]{guena2011doppler}
\bibinfo{author}{\bibfnamefont{J.}~\bibnamefont{Gu\'ena}}, \bibinfo{author}{\bibfnamefont{R.}~\bibnamefont{Li}}, \bibinfo{author}{\bibfnamefont{K.}~\bibnamefont{Gibble}}, \bibinfo{author}{\bibfnamefont{S.}~\bibnamefont{Bize}}, \bibnamefont{and} \bibinfo{author}{\bibfnamefont{A.}~\bibnamefont{Clairon}}, \bibinfo{journal}{Phys. Rev. Lett.} \textbf{\bibinfo{volume}{106}}, \bibinfo{pages}{130801} (\bibinfo{year}{2011}), \urlprefix\url{https://link.aps.org/doi/10.1103/PhysRevLett.106.130801}.

\bibitem[{\citenamefont{Wineland}(2013)}]{wineland2013nobel}
\bibinfo{author}{\bibfnamefont{D.~J.} \bibnamefont{Wineland}}, \bibinfo{journal}{Rev. Mod. Phys.} \textbf{\bibinfo{volume}{85}}, \bibinfo{pages}{1103} (\bibinfo{year}{2013}), \urlprefix\url{https://link.aps.org/doi/10.1103/RevModPhys.85.1103}.

\bibitem[{\citenamefont{Zhang et~al.}(2017)\citenamefont{Zhang, Deng, Luo, and Lu}}]{zhang2017direct}
\bibinfo{author}{\bibfnamefont{J.}~\bibnamefont{Zhang}}, \bibinfo{author}{\bibfnamefont{K.}~\bibnamefont{Deng}}, \bibinfo{author}{\bibfnamefont{J.}~\bibnamefont{Luo}}, \bibnamefont{and} \bibinfo{author}{\bibfnamefont{Z.-H.} \bibnamefont{Lu}}, \bibinfo{journal}{Chinese Physics Letters} \textbf{\bibinfo{volume}{34}}, \bibinfo{pages}{050601} (\bibinfo{year}{2017}), \urlprefix\url{https://iopscience.iop.org/article/10.1088/0256-307X/34/5/050601/meta}.

\bibitem[{\citenamefont{Huang et~al.}(2022)\citenamefont{Huang, Zhang, Zeng, Hao, Ma, Zhang, Guan, Chen, Wang, and Gao}}]{huang2022liquid}
\bibinfo{author}{\bibfnamefont{Y.}~\bibnamefont{Huang}}, \bibinfo{author}{\bibfnamefont{B.}~\bibnamefont{Zhang}}, \bibinfo{author}{\bibfnamefont{M.}~\bibnamefont{Zeng}}, \bibinfo{author}{\bibfnamefont{Y.}~\bibnamefont{Hao}}, \bibinfo{author}{\bibfnamefont{Z.}~\bibnamefont{Ma}}, \bibinfo{author}{\bibfnamefont{H.}~\bibnamefont{Zhang}}, \bibinfo{author}{\bibfnamefont{H.}~\bibnamefont{Guan}}, \bibinfo{author}{\bibfnamefont{Z.}~\bibnamefont{Chen}}, \bibinfo{author}{\bibfnamefont{M.}~\bibnamefont{Wang}}, \bibnamefont{and} \bibinfo{author}{\bibfnamefont{K.}~\bibnamefont{Gao}}, \bibinfo{journal}{Phys. Rev. Applied} \textbf{\bibinfo{volume}{17}}, \bibinfo{pages}{034041} (\bibinfo{year}{2022}).

\bibitem[{\citenamefont{Phillips}(1998)}]{phillips1998laser}
\bibinfo{author}{\bibfnamefont{W.~D.} \bibnamefont{Phillips}}, \bibinfo{journal}{Rev. Mod. Phys.} \textbf{\bibinfo{volume}{70}}, \bibinfo{pages}{721} (\bibinfo{year}{1998}), \urlprefix\url{https://link.aps.org/doi/10.1103/RevModPhys.70.721}.

\end{thebibliography}
\end{document}